\documentclass[twocolumn,preprintnumbers,elsart]{revtex4}

\usepackage{amssymb}
\usepackage{threeparttable}
\usepackage{amsmath}
\usepackage{graphicx}
\usepackage{dcolumn}
\usepackage[center]{subfigure}
\usepackage{array}
\usepackage{booktabs}
\usepackage{multirow}
\usepackage{float}
\usepackage{color}
\usepackage{epstopdf}
\usepackage{hyperref}
\usepackage{mathtools} 

\begin{document}

\title{Chiral solitons in quadratic quasi-phase-matched photonic crystals}

\author{Yuxin Guo$^{1}$}
\author{Xuening Wang$^{1}$}
\author{Zhiwei Fan$^{2}$}
\author{Zhaopin Chen$^{3}$}
\author{Qiuyi Ning$^{1,4}$}
\author{Hexiang He$^{1,4}$}
\email{sysuhhx@163.com}
\author{Wei Pang$^{5}$}
\author{Li Zhang$^{1,4}$}
\email{zhangli4102@126.com}
\author{Yongyao Li$^{1,4}$}

\affiliation{$^{1}$School of Physics and Optoelectronic Engineering, Foshan University, Foshan 528000, China\\
$^{2}$School of Engineering, Newcastle University, Newcastle upon Tyne NE1 7RU, UK\\
$^{3}$Physics Department and Solid-State Institute, Technion, Haifa 32000, Israel\\
$^{4}$Guangdong-Hong Kong-Macao Joint Laboratory for Intelligent Micro-Nano Optoelectronic Technology, School of Physics and Optoelectronic Engineering, Foshan University, Foshan 528000, China\\
$^{5}$Department of Experiment Teaching, Guangdong University of Technology, Guangzhou 510006, China}

\date{\today}

\begin{abstract}
We introduce a quasi-phase-matched technique in quadratic nonlinear crystals, constructing an artificial gauge field by changing the inclination angle of stripes, which is realized by the positive and negative polarization directions of nonlinear susceptibility along the crystal. Unlike the artificial gauge field constructed through linear coupling in other settings, the gauge field in this system is realized by nonlinear coupling. We demonstrate that this gauge field can generate stable chiral solitons with chiral energy flow rotating around the solitons. In contrast to conventional chiral currents generated with the same specie or frequency, the chiral currents in the present system are formed by mutual coupling between fundamental frequency and second harmonic components. We derive the semi-analytical solution for the chiral energy flow in this system. It is found that there exists an optimal inclination angle that can maximize the chiral energy flow under different parameters, and this optimal inclination shows a positive correlation with the power and detuning. The mobility and collisions of the chiral solitons are also discussed. The results show that chiral solitons move in response to kicking and undergo fully elastic collisions with each other. In addition, the possibility of experimentally generating chiral solitons and chiral currents is outlined.
\end{abstract}

\maketitle


\section{Introduction}
The applications of chiral solitons are widely distributed in the fields of high-energy physics \cite{JS2004,AJ2020}, condensed matter physics \cite{ME2013,JS2023,MM2010}, optics \cite{VI2003,YS2021}, and materials science \cite{CZ2021,SR2017}, etc., and they have become an important research object because of their topological properties and stabilities \cite{RD2019,RD2018,MJ2015}. The existence of chiral solitons relies on chiral symmetry, i.e., the different manifestations of left- and right-spin components under a specific symmetry group, which is characterized by non-reciprocity and unidirectionality \cite{UA1996,EH1998}. In nonlinear optics, chiral solitons can be applied to fiber optic communication, optical transmission and other fields \cite{SS2008,FJ2010}. It has been shown that the coupling of nonlinear excitation with the system background creates a competitive relationship. This interaction, between current nonlinearity from the canonical potential and mean-field atomic forces, enables the generation of chiral solitons \cite{RG2022,KF2012,WN2024}. Therefore, finding a suitable canonical potential or constructing an equivalent artificial gauge field becomes the key to prepare chiral solitons. The synthetic gauge theory developed in recent years enables coupling between photons and synthetic magnetic fields in nonmagnetic photonic systems \cite{RO2011,SL2013,FL2015,NS2016,SM2023}. In past researches, artificial gauge fields in optical lattices have been constructed by laser-assisted tunnelling \cite{MA2014,AC2014,DH2014}, applying voltages to change the birefringence coefficient of the crystals \cite{GL2023,GL2024}, and so on. In all these past studies, the construction of artificial gauge fields was achieved by linear couplings. In this work, we adopt a new approach to construct artificial gauge fields in nonlinear couplings and through the quasi-phase-matched (QPM) technique in nonlinear crystals, where the modulation phase of the QPM can be shifted by changing the inclination of the positive and negative polarization directions of the nonlinear susceptibility in the crystals, which is related to the $X$-direction, and this phase modulation is equivalent to the introduction of an artificial magnetic field. Note that similar consideration of introducing the equivalent magnetic fields through the technique of QPM had been applied to produced optical Stern-Gerlach effect \cite{OY2022,AKa2018,Aviv2018} and the adiabatic geometric phase in the nonlinear frequency conversion \cite{AK2022,AK2018,AK2019,YL2020,FZ2021,YL2021}.

Quadratic ($\chi^{(2)}$) nonlinear crystals with strong and fast nonlinear optical response have been proven to be favorable platforms for studying soliton transport \cite{CE2000,VE1999,BZ2016}, which can create ideal conditions for optical soliton transport due to their unique nonlinear effects \cite{JY2003,JY2004,JYI2004,YS2008}. In $\chi^{(2)}$ nonlinear crystals, different wavelengths of light waves have different refractive indices, resulting in mismatch of their wave vectors, and thus phase matching cannot be realized naturally \cite{HS2023,MM1992,JL2022}, which seriously affects the frequency conversion efficiency, and the QPM technique can solve this problem. The QPM technique was first proposed by Bloembergen in 1962 has developed into a mature technology \cite{FZ2023,AB2010,DZ2008,KM1994,ZF2025,CK2024}. In QPM crystals, the alternating positive and negative polarization domains are achieved through ferroelectric domain inversion \cite{BZ2025}, enabling wave vector mismatch compensation via spatial modulation of the quadratic nonlinear susceptibility $\chi^{(2)}$ \cite{AP2004,SZ1997,LT2011,AA2010}, and the linear refractive index remains invariant \cite{YZ2021}. The QPM technique can significantly improve the frequency conversion efficiency of nonlinear optical processes \cite{CB2002,MS2009,AB2008,VB1998}, and it has now been widely used in various nonlinear optical frequency conversion scenarios \cite{YG2014,SL2023,YT2024}.
\begin{figure}[htbp]
\centering
{\includegraphics[scale=0.24]{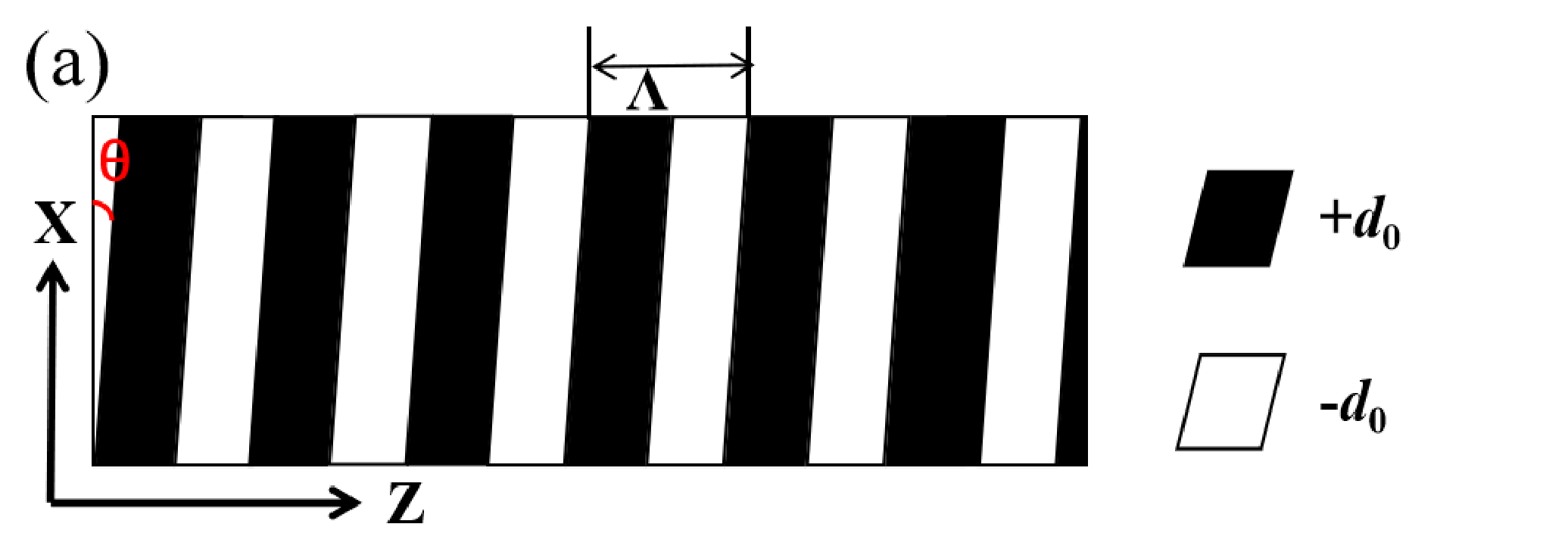}}
\caption{Schematic diagram of a nonlinear photonic crystal cross-section in the transmission direction. White areas represent positive polarization directions and black areas represent negative polarization directions.}\label{trans}
\end{figure}

In this paper, we construct artificial gauge fields by introducing the QPM technique to the nonlinear crytals, which varies the inclination of the positive and negative polarization directions of the crystals to generate a spatially correlated phase, and find that chiral solitons carrying currents can be generated in the interaction of an equivalent magnetic field with a coupled wave. Unlike previous chiral currents by the same specie or same frequency \cite{RW2014,QY2018}, the chiral current in the present system is co-constituted by fundamental frequency (FF) and second harmonic (SH) waves coupled to each other. We find that under different parameters, the modulation of the crystal has an angle that maximises the energy current, which we call the optimal inclination (OI). The transport properties of chiral solitons and the effect of parameters on OI are illustrated through the paper. In this work, we use the imaginary time propagation (ITP) method to converge the solution of chiral soliton \cite{FY2008,SS2006}, and use the direct simulations as well as the Bogoliubov¨Cde Gennes (BdG) equations to verify the stability of the soliton solution \cite{KK2017,JF2003}. It is found that the chiral soliton can be stably transmitted in a $\chi^{(2)}$ nonlinear photonic crystal, and the OI is related to the power and detuning. In addition to the numerical simulations, we have derived the semi-analytical solution for chiral energy flow, and the results obtained from the calculations are consistent with the trend of the numerical simulations. The paper is structured as follow: The model and the analysis are illustrated in Sec. II; the numerical results and discussions are presented in Sec. III. Mobilities and collisions of the chiral solitons are also briefly discussed in Sec. IV; estimations of the creation chiral soliton in experiment is briefly outlined in Sec. V; the paper is concluded in Sec. VI.
\section{The model}

We use QPM technique to modulate the solitons in the transmission direction $Z$. The equations for the evolution of the FF and SH field amplitudes with propagation distance $Z$ are:
\begin{eqnarray}
i\partial_{Z}A_{1}&=&-\frac{1}{2k_{1}}\partial_{X}^2A_{1}-\frac{2d(Z)\omega_{1}}{cn_{1}}A_{1}^{*}A_{2}e^{-i\Delta k Z}\label{dynamiceq1},\\
i\partial_{Z}A_{2}&=&-\frac{1}{2k_{2}}\partial_{X}^2A_{2}-\frac{d(Z)\omega_{2}}{cn_{2}}A_{1}^{2}e^{i\Delta k Z}\label{dynamiceq2},
\end{eqnarray}
where $A_{1, 2}$ are the slow-varying amplitudes of the FF and SH waves, respectively; $k_{1, 2}$ , $\omega_{1, 2}$ ($\omega_{2}=2\omega_{1}$), and $n_{1, 2}$ are the wave vector, angular frequency and refractive index in a nonlinear crystal for FF and SH waves respectively, $c$ is the speed of light propagation in a vacuum, and $\Delta k=2k_{1}-k_{2}$ [$k_{j}=2\pi n_{j}/\lambda_{j}(j=1,2)$, $\lambda_{1,2}$ are the wavelengths of FF and SH waves, respectively] is the phase mismatch factor.
\begin{eqnarray}
d(Z)=d_0\mathrm{sgn}\left[\cos\left(2\pi Z/\Lambda +\varphi_{d}\right)\right], \label{Dz-real}
\end{eqnarray}
is used to describe the modulation coefficient of the nonlinear susceptibility in crystals,where $\Lambda$ is the inversion period of the QPM crystal. It can be written as a Fourier series expansion of the form \cite{HS2014,AK2022}:
\begin{eqnarray}
d(Z)=d_{0}\sum^{ }_{m\neq0}\frac{2}{\pi m}\sin(\frac{\pi m}{2})\exp[im(\frac{2\pi }{\Lambda}Z+\varphi_{d})],\label{Dz2}
\end{eqnarray}
where $d_{0}$ is the $\chi^{(2)}$ polarizability tensor. Note the periodical poling of $d(Z)$ do not change the linear refractive index of crystals, its linear dispersion relation curve is a parabola, hence no linear bandgap is created by this modulation \cite{YZ2021}. As shown in
Fig. \ref{trans}, in Eq. (\ref{Dz2}), the adjacent positive polarized part and negative polarized part consist a modulation cycle whose period is $\Lambda$. The polarized modulation duty cycle is 1/2. By tilting the crystal in the direction of positive and negative polarization, the square wave function is shifted along $X$, resulting in the phase $\varphi_{d}$ as a function of $X$. Since all the stripes of the positive and negative polarization are tilted by an angle $\theta$ [see in Fig. \ref{trans}], this configuration induces $\varphi_{d}$ that can be expressed as a linear function of $X$, given by:
\begin{eqnarray}
\varphi_{d}=\frac{2\pi}{\Lambda}\Delta Z=\frac{2\pi}{\Lambda}\tan\theta X.\label{d}
\end{eqnarray}

We assumed that the terms of $m=\pm1$ play the dominant role in the compensation for the phase mismatch. See Appendix B for more details. By applying rescaling \cite{YL2020,FZ2021}:
\begin{eqnarray}
&&u={A}_{1}\sqrt{\frac{n_{1}}{\omega_{1}I_{0}}} \exp[i(\Delta k-\frac{2\pi}{\Lambda})Z],\notag \\
&&v={A}_{2}\sqrt{\frac{n_{2}}{\omega_{2}I_{0}}} \exp[i(\Delta k-\frac{2\pi}{\Lambda})Z],\notag \\
&&I_{0}=(\frac{n_{1}}{\omega_{1}}+\frac{n_{2}}{\omega_{2}})|A_{0}|^{2},\quad
z^{-1}_{d}=\frac{2d_{0}\omega_{1}}{\pi cn_{1}}\sqrt{\frac{\omega_{2}}{n _{2}}I_{0}},\notag \\
&&z=Z/z_{d},\quad x=X\sqrt{k_{1}/z_{d}},\notag \\
&&k_{1}/k_{2}\approx2,\quad n_{1}\approx n_{2}=n,\quad \alpha=\frac{2\pi}{\Lambda}\tan\theta\sqrt{\frac{z_{d}}{k_{1}}},\label{units}
\end{eqnarray}
two dimensionless equations are then obtained:
\begin{eqnarray}
&&i\partial_{z}u=-\frac{1}{2}\ \frac{\partial^{2}u}{\partial x^{2}}\ - \Omega u- 2e^{i\alpha x}u^{*}v, \label{units1}\\
&&i\partial_{z}v=-\frac{1}{4}\ \frac{\partial^{2}v}{\partial x^{2}}\ - \Omega v- e^{-i\alpha x}u^{2}, \label{units2}
\end{eqnarray}
where $A_{0}$ is a characteristic amplitude of the electromagnetic field, $\alpha$ is a parameter related to the tilt angle $\theta$, and
\begin{eqnarray}
\Omega=z_{d}(\Delta k-\frac{2\pi}{\Lambda}), \label{detuning}
\end{eqnarray}
is the detuning. Therefore, when $|\Delta k|=2\pi/\Lambda$, the detuning is 0 and phase matching can be achieved \cite{RW2009}. If we want the detuning not to be 0, we only need to change the wavelength of the incident wave so that it is shifted from 1064nm. See Appendix A for more details.

The total Hamiltonian and power are \cite{WM2009,EA2019,GP2013}:
\begin{eqnarray}
\begin{aligned}
&H=\int\frac{1}{2}(|\partial_{x}u|^{2}+\frac{1}{2}|\partial_{x}v|^{2}) \\ &\quad\quad-\Omega(|u|^{2}+|v|^{2})-(u^{*2}v+c.c)\mathrm{d}x,\label{h1}
\end{aligned}
\end{eqnarray}
\begin{eqnarray}
P=\int{\left(|u|^{2}+2|v|^{2}\right)}\mathrm{d}x=P_{u}+2P_{v}.\label{power}
\end{eqnarray}

The steady-state soliton solution of the dipole soliton is expressed in terms of the propagation constant $\beta$ of the two waves as:
\begin{eqnarray}
&&u(z,x)=\psi_{u}(x)e^{i\beta_{1} z}, \label{uzx}\\
&&v(z,x)=\psi_{v}(x)e^{i\beta_{2} z}, \label{vzx}
\end{eqnarray}
where $\psi_{u,v}(x)$ represent the stationary solution of the FF and SH waves, with propagation constants, which are real numbers, $\beta_{1}$ and $\beta_{2}$, respectively. Substitute $u(z,x)$ and $v(z,x)$ into Eqs. (\ref{units1},\ref{units2}), the steady-state equations are:
\begin{eqnarray}
&&\beta_{1}\psi_{u}=-\frac{1}{2}\ \frac{\partial^{2}\psi_{u}}{\partial x^{2}}\ -\Omega \psi_{u}-2e^{i\alpha x}\psi_{u}^{*}\psi_{v}, \label{beta1}\\
&&\beta_{2}\psi_{v}=-\frac{1}{4}\ \frac{\partial^{2}\psi_{v}}{\partial x^{2}}\ -\Omega \psi_{v}-e^{-i\alpha x}\psi_{u}^{2}, \label{beta2}
\end{eqnarray}
the following relationship exists between the two propagation constants:
\begin{eqnarray}
&&\beta_{2}=2\beta_{1}. \label{pc4}
\end{eqnarray}
The stability analysis is conducted by introducing perturbations to the solutions as follows:
\begin{eqnarray}
\{u,v\}=[\psi_{\{u,v\}}+p_{\{u,v\}}e^{-i\Gamma z}+q_{\{u,v\}}e^{i\Gamma z}]e^{i\beta_{\{1,2\}}z}, \label{u,v}
\end{eqnarray}
where $p,q$ are perturbation terms and $\Gamma$ is the eigenvalue. Substituting the perturbation wave function in the form as above into Eqs. (\ref{units1},\ref{units2}) for the Bogoliubov-de Gennes (BdG) analysis linearized the eigenvalue matrix as follows:
\begin{equation}
\begin{bmatrix}
\hat{M}_1 & \hat{M}_2 & \hat{M}_3 & 0 \\
-\hat{M}^{*}_2 & -\hat{M}^{*}_1 & 0 & -\hat{M}^{*}_3 \\
\hat{M}_4 & 0 & \hat{M}_5 & 0 \\
0 & -\hat{M}^{*}_4 & 0 & -\hat{M}^{*}_5
\end{bmatrix}
\begin{bmatrix}
p_u \\
q_u \\
p_v \\
q_v
\end{bmatrix}
= \Gamma
\begin{bmatrix}
p_u \\
q_u \\
p_v \\
q_v
\end{bmatrix}
, \label{zeta}
\end{equation}
with operators:
\begin{eqnarray}
&&\hat{M}_1=-\frac{1}{2}\partial_{x^{2}}-\Omega+\beta_{1},\notag \\
&&\hat{M}_2=-2e^{i\alpha x}v,\quad \hat{M}_3=-2e^{i\alpha x}u^{*},\quad \hat{M}_4=-2e^{-i\alpha x}u,\notag \\
&&\hat{M}_5=-\frac{1}{4}\partial_{x^{2}}-\Omega+\beta_{2},\label{M}
\end{eqnarray}
when all eigenvalues $\Gamma$ are real, the stationary solution is stable.

When $\alpha=0$, $\psi_{u}$ and $\psi_{v}$ are real functions. When $\alpha\neq0$, we assume:
\begin{eqnarray}
&&\psi_{u}(x)=\phi_{u}e^{i\eta_{u} x}, \label{psiu}\\
&&\psi_{v}(x)=\phi_{v}e^{i\eta_{v} x}, \label{psiv}
\end{eqnarray}
where $\phi_{u}(x)$ and $\phi_{v}(x)$ represent the modulus of complex wave function of the FF and SH waves, $\eta_{u,v}$ are two real numbers. According to the phase matching relationship between the two sides of the equality sign of the two equations, we can get:
\begin{eqnarray}
&&2\eta_{u}-\eta_{v}=\alpha. \label{psiuv}
\end{eqnarray}
We assume that the power density is:
\begin{eqnarray}
&&\rho=|\psi_{u}|^{2}+2|\psi_{v}|^{2}, \label{rho}
\end{eqnarray}
from the energy flow conservation equation $\partial_{z}\rho+\nabla\cdot j=0$, a partial derivation of Eq. (\ref{rho}) in the $Z$ direction is obtained:
\begin{eqnarray}
\begin{aligned}
\partial_{z}\rho&=\frac{i}{2}\partial_{x}(\psi_{u}^{*}\partial_{x}\psi_{u}-\psi_{u}\partial_{x}\psi_{u}^{*})\\
&\quad+\frac{i}{2}\partial_{x}(\psi_{v}^{*}\partial_{x}\psi_{v}-\psi_{v}\partial_{x}\psi_{v}^{*})-8\mathrm{Im}(e^{i\alpha x}\psi_{u}^{*2}\psi_{v}),\label{rho2}
\end{aligned}
\end{eqnarray}
by definition, the probability flow densities of the FF and the SH waves are \cite{JS2023}:
\begin{eqnarray}
&&j_{u}=-\frac{i}{2}(\psi_{u}^{*}\partial_{x}\psi_{u}-c.c), \label{less1}\\
&&j_{v}=-\frac{i}{4}(\psi_{v}^{*}\partial_{x}\psi_{v}-c.c), \label{less2}
\end{eqnarray}
thus, Eq. (\ref{rho2}) can be written as£º
\begin{eqnarray}
\partial_{z}\rho+\partial_{x}(j_{u}+2j_{v})=-8\mathrm{Im}(e^{i\alpha x}\psi_{u}^{*2}\psi_{v}),\label{rho3}
\end{eqnarray}
since $P=\int_{-\infty}^{\infty}\rho \mathrm{d}x$ is a conserved quantity, $\partial_{z}P\equiv0$. Integrate Eq. (\ref{rho3}) in the $X$ direction:
\begin{eqnarray}
\begin{aligned}
j_{u}+2j_{v}&=\int_{-\infty}^{\infty}-8\mathrm{Im}(e^{i\alpha x}\psi_{u}^{*2}\psi_{v})\mathrm{d}x\\
&=\int_{-\infty}^{\infty}-8i\sin[(\alpha-2\eta_{u}+\eta_{v})x]|\phi_{u}|^2|\phi_{v}|\mathrm{d}x.\label{rho4}
\end{aligned}
\end{eqnarray}
According to Eq. (\ref{psiuv}), $\alpha-2\eta_{u}+\eta_{v}\equiv0$, yields,
\begin{eqnarray}
j_{u}+2j_{v}=\eta_{u}|\phi_{u}|^2+\eta_{v}|\phi_{v}|^2=0. \label{juv}
\end{eqnarray}
Here, we have adopted $j_{u}=\eta_{u}|\phi_{u}|^2$ and $j_{v}=\eta_{v}|\phi_{v}|^2/2$ by substituting Eqs. (\ref{psiu},\ref{psiv}) into Eqs. (\ref{less1},\ref{less2}). Taking into account with Eq. (\ref{psiuv}), one can yield,
\begin{eqnarray}
&&\eta_{u}=\frac{\alpha P_{v}}{2P}, \label{etau}\\
&&\eta_{v}=-\frac{\alpha P_{u}}{P}, \label{etav}
\end{eqnarray}
at the same time, the energy flow density of the FF and the SH waves are:
\begin{eqnarray}
&&J_{u}=-\beta_{1}j_{u}=-\beta_{1}\frac{\alpha P_{v}}{2P}|\phi_{u}|^2, \label{Ju}\\
&&J_{v}=-\beta_{2}j_{v}=\beta_{1}\frac{\alpha P_{u}}{2P}|\phi_{v}|^2, \label{Jv}
\end{eqnarray}
here, we replace $\beta_{2}$ in Eq. (\ref{Jv}) with $2\beta_{1}$ (see Eq. (\ref{pc4})). Further, the energy flow of the two waves are:
\begin{eqnarray}
&&S_{u}=\int J_{u} \mathrm{d}x=-\beta_{1}\frac{\alpha P_{u}P_{v}}{2P}, \label{Su}\\
&&S_{v}=\int J_{v} \mathrm{d}x=\beta_{1}\frac{\alpha P_{u}P_{v}}{2P}, \label{Sv}
\end{eqnarray}
from Eqs. (\ref{Su},\ref{Sv}), we know that $S_{u}$ and $S_{v}$ are equal in magnitude and opposite in direction. When $S_{u}<0$, we define its corresponding energy flow direction to be right-to-left, and correspondingly, the energy flow direction of the SH wave is left-to-right at this point. When $\alpha>0$, the direction of the energy flow conforms to the left-handed helix rule; when $\alpha<0$, the direction of the energy flow conforms to the right-handed helix rule. Here, we define the chiral energy flow in the system as:
\begin{eqnarray}
&&S_{c}=S_{u}-S_{v}=-\alpha\frac{\beta_{1} P_{u}P_{v}}{P}.\label{sc}
\end{eqnarray}
{Thus, through the above semi-analytical process, we have obtained the expression for the chiral energy flow, where $S_c$ is a quantity determined by the input parameters. $S_c$ describes the asymmetry and chiral characteristics of the energy flow in the system, if $\alpha=0$, then $S_c=0$, it means that the energy flow of the FF wave and the SH wave are equal, and there is no chirality in the system; if $\alpha\neq0$, then $S_c\neq0$, it means that the two are asymmetric, and there is a chiral characteristic in the system.

\begin{figure}[t]
\centering
{\includegraphics[scale=0.19]{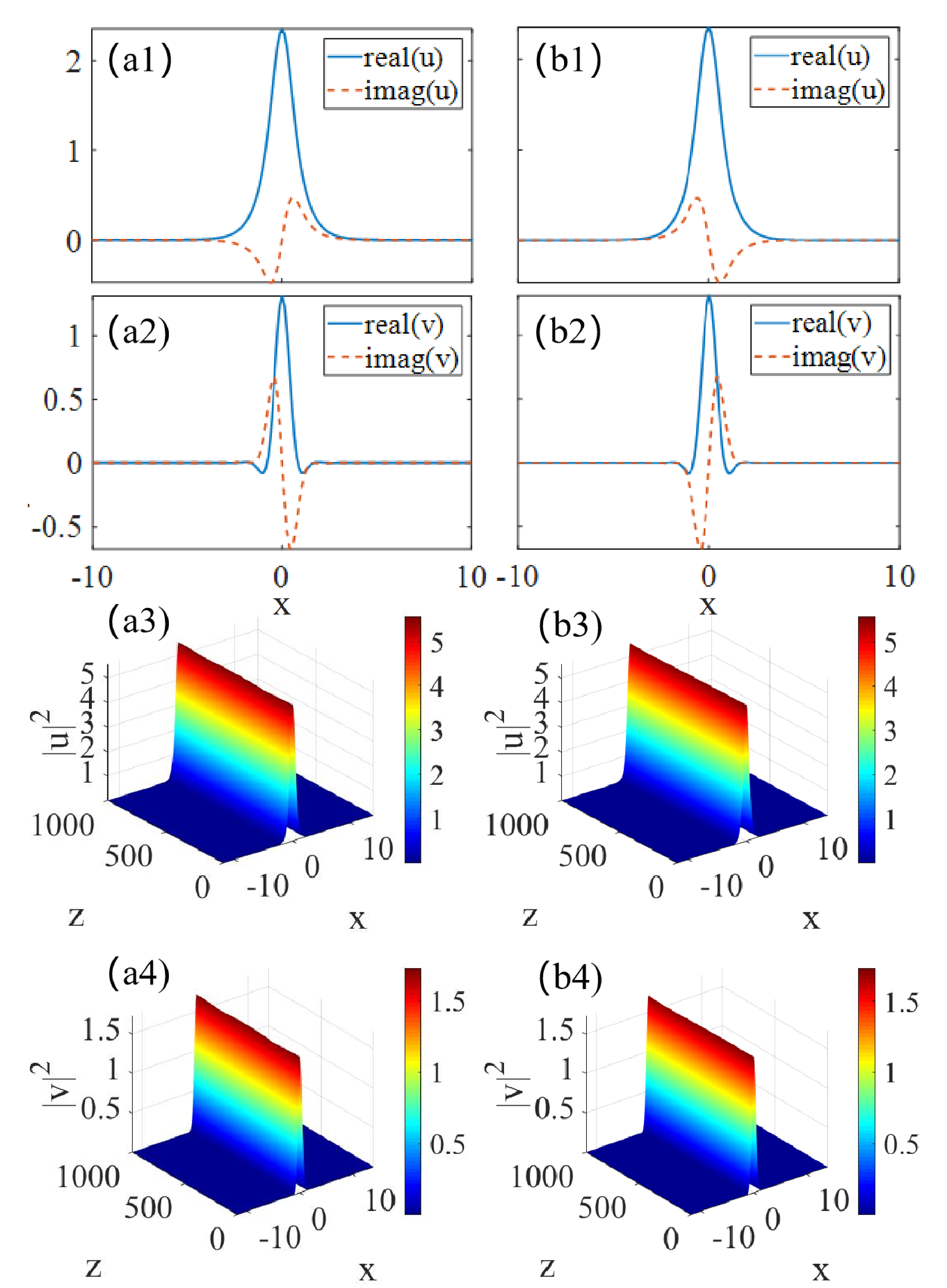}}
\caption{(a1-a2)Amplitude of the chiral soliton solution with $(P, \Omega, \alpha)=(10,0,4)$, where $u$ is the FF component, $v$ is the SH component. The solid blue curve represents the real part and the dashed red line represents the imaginary part of amplitude. (b1-b2) The amplitude of the solitons solution with ($P$, $\Omega$, $\alpha$) = (10, 0, -4). (a3,a4,b3,b4) Direct simulation of chiral solitons (perturbed by 1\% noises), which shown by the intensity for the FF (a3,b3) and SH (a4,b4) components, respectively, at ($P$, $\Omega$, $\alpha$) = (10, 0, 4) (a3,a4) and (10,0,-4) (b3,b4). }\label{ITPfig}
\end{figure}

\section{Numerical Results}

We find that an artificial gauge field can be constructed by introducing a QPM technique in a nonlinear crystal, and the coupled wave interacts with the equivalent magnetic field to produce chiral solitons. Figs. \ref{ITPfig}(a1,a2) display the soliton solutions for the FF and SH waves, obtained through convergence under the ITP method at $\alpha>0$. Here, the blue and red dashed curves represent the real and imaginary parts, respectively. Figs. \ref{ITPfig}(b1,b2) show the corresponding case for $\alpha<0$. Direct simulations of these chiral solitons, presented in Figs. \ref{ITPfig}(c1,c2), confirm their stability. In these figures, we can see that the soliton solution exhibit a form similar to the expressions in Eqs. (\ref{psiu}, \ref{psiv}), where $\psi_{u,v}$ adopt a Gaussian-like profile. In this configuration, the exponential term, $\exp(i\eta_{u,v}x)$ can be expanded using Euler's formula, resulting in the real and imaginary parts of the soliton. Fig. \ref{ITPphase} reveals that the slopes of the phase curves indicate opposite signs for ${\eta_{u}}$ and ${\eta_{v}}$, consistent with the prediction in Eqs. (\ref{etau},\ref{etav}).

\begin{figure}[htbp]
\centering
{\includegraphics[scale=0.19]{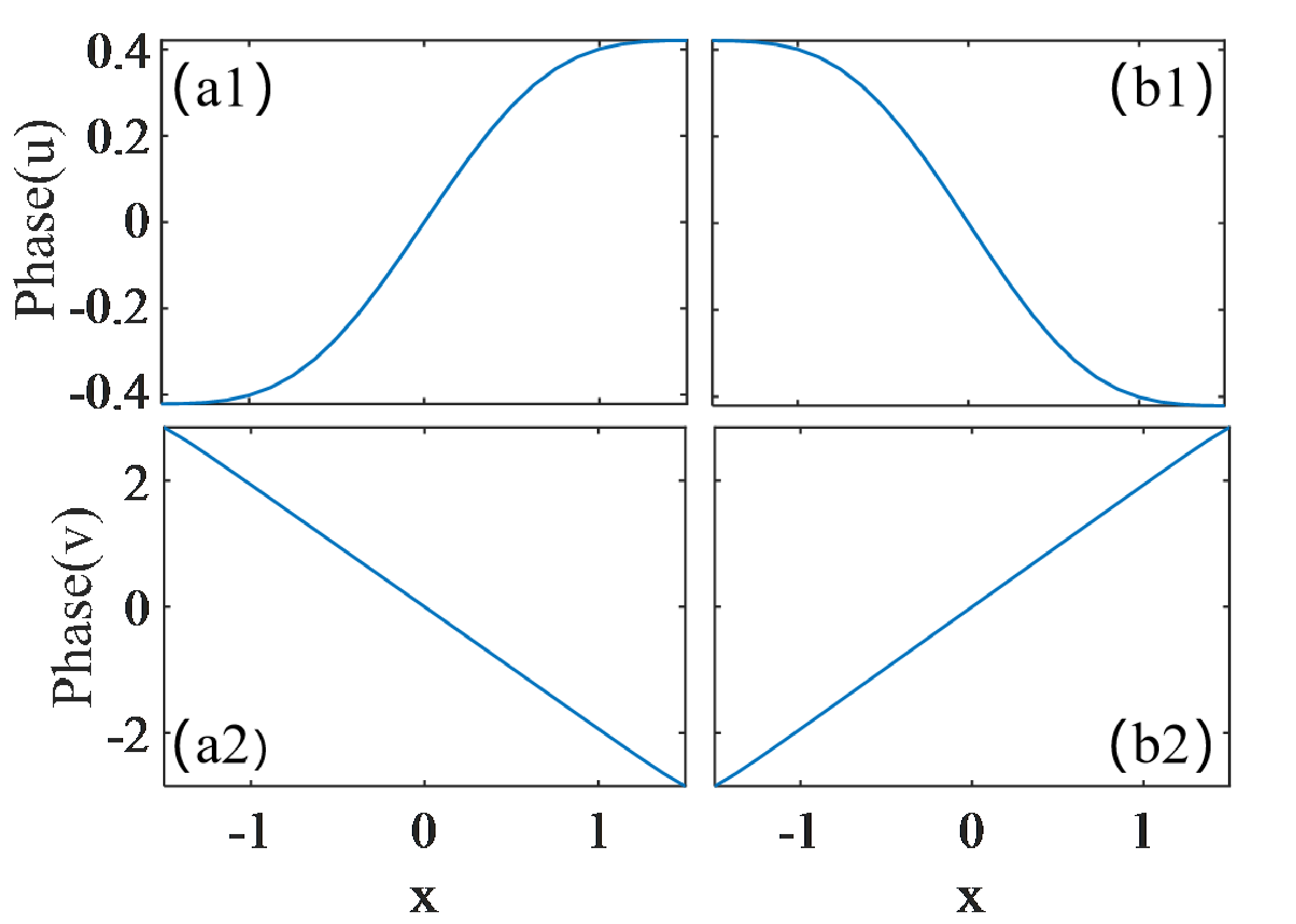}}
\caption{The phase distribution of the solitons
corresponding to Figs. 2(a1-b2), respectively.}\label{ITPphase}
\end{figure}

Fig. \ref{stable} shows the stabilization regions of chiral solitons in the plane of detuning $\Omega$ and tilt angle $\alpha$ with different total power $P$, which we obtained by direct transmission. At the same time, we have verified the stability of the soliton solutions in the above stabilized regions using the BDG method, and their eigenvalues are less than 0.01. Here, we have selected three powers of $P=\{5,15,20\}$. In the green region, the soliton is stable at all three powers; in the red region, the soliton is stable only at $P=10$ and $P=20$; in the blue region, the soliton is stable only at $P=20$. The results show that the solitons are stable within the solution context. However, when $\Omega>0$, as detuning increases, the soliton radius gradually increases. When $\Omega<0$, there exists a threshold above which a soliton can be formed only if the $P$ is higher than this threshold, otherwise there is no soliton solution, and this threshold increases as the $\Omega$ is less than zero  \cite{RA1997,AB2002}.

 \begin{figure}[htbp]
\centering
{\includegraphics[scale=0.3]{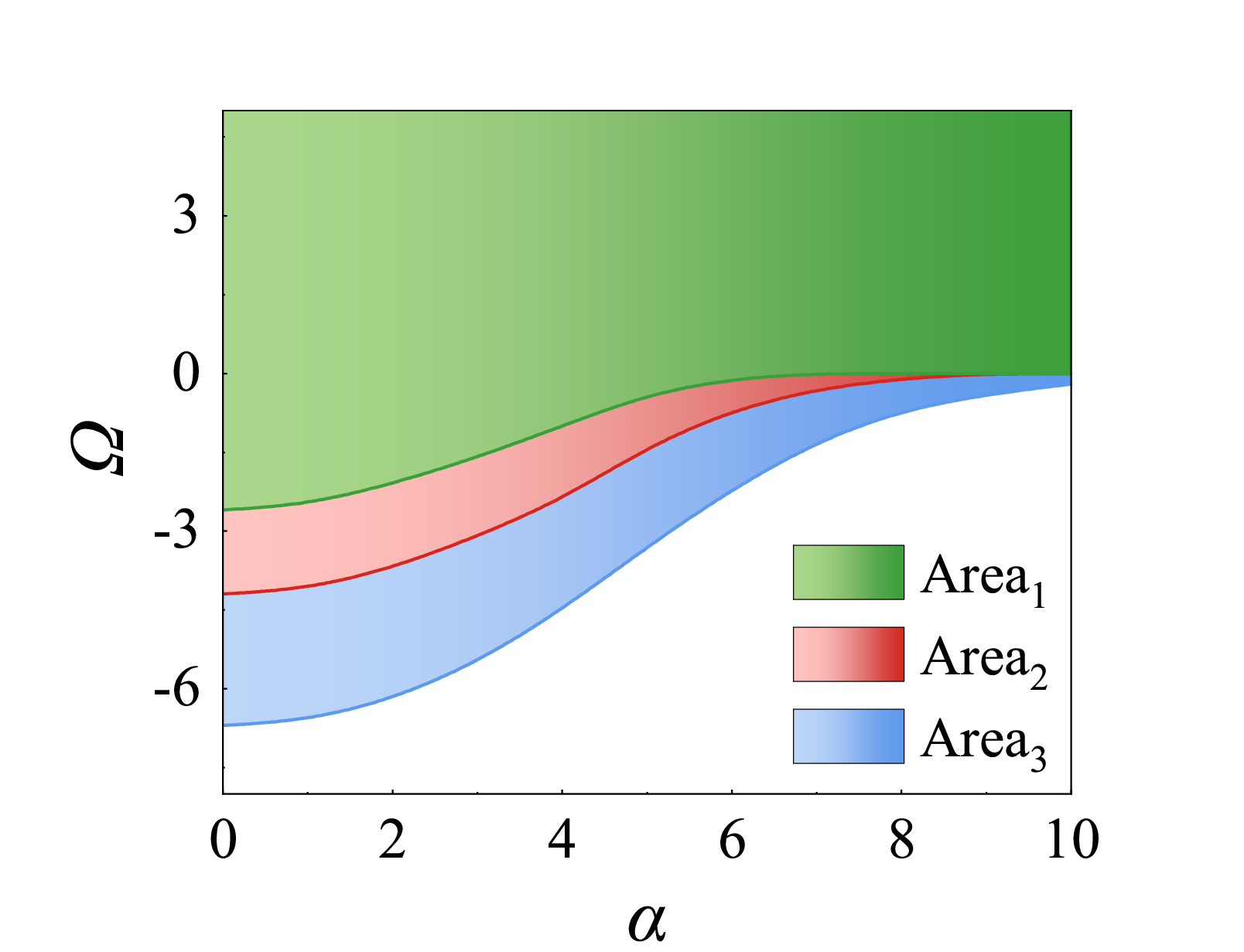}}
\caption{The stabilization regions of chiral solitons in the plane of detuning $\Omega$ and tilt angle $\alpha$ with different total power $P$. Stabilization regions are color-coded with $\mathrm{Area_{1}}$ (green) covering $P=\{5,10,20\}$, $\mathrm{Area_{2}}$ (red) covering $P=\{10,20\}$, $\mathrm{Area_{3}}$ (blue) covering $P=20$.}\label{stable}
\end{figure}

To investigate the propagation properties of chiral solitons, we have numerically analyzed the relationship between the propagation constants of chiral solitons and each parameter. Fig. \ref{beta}(a) shows that the propagation constant $\beta_{1}$ increases with the total power $P$ for the angular parameter $\alpha (\alpha = 2, 4, 6)$, i.e., the higher the $P$ is, the more stable the soliton transmission is, which satisfies the Vakhitov-Kolokolov criterion, i.e., the slope $d\beta/dP>0$, which is a necessary condition for the formation of stable solitons in a self-focusing medium. Moreover, at the same $P$, $\beta^{(\alpha=2)}_{1}> \beta^{(\alpha=4)}_{1}> \beta^{(\alpha=6)}_{1}$ exists. In Fig. \ref{beta}(b), the $\beta_{1}$ shows a negative correlation with the absolute value of $\alpha$, which mirrors the results of Fig. \ref{beta}(a). Fig. \ref{beta}(c) shows that the ratio of the power of the FF wave to the total power, $P_{u}/P$, increases as $|\alpha|$ increases. From Fig. \ref{beta}(c) we can conclude that the larger $|\alpha|$ is, the higher the component of the fundamental frequency wave and the lower the component of the second harmonic. And Fig. 5(d) shows that the product of the power of the FF wave and the SH wave, $P_{u}P_{v}$, decreases as $|\alpha|$ increases, which is consistent with the variation of $\beta_{1}$ with $|\alpha|$.

\begin{figure}[b]
\centering
{\includegraphics[scale=0.298]{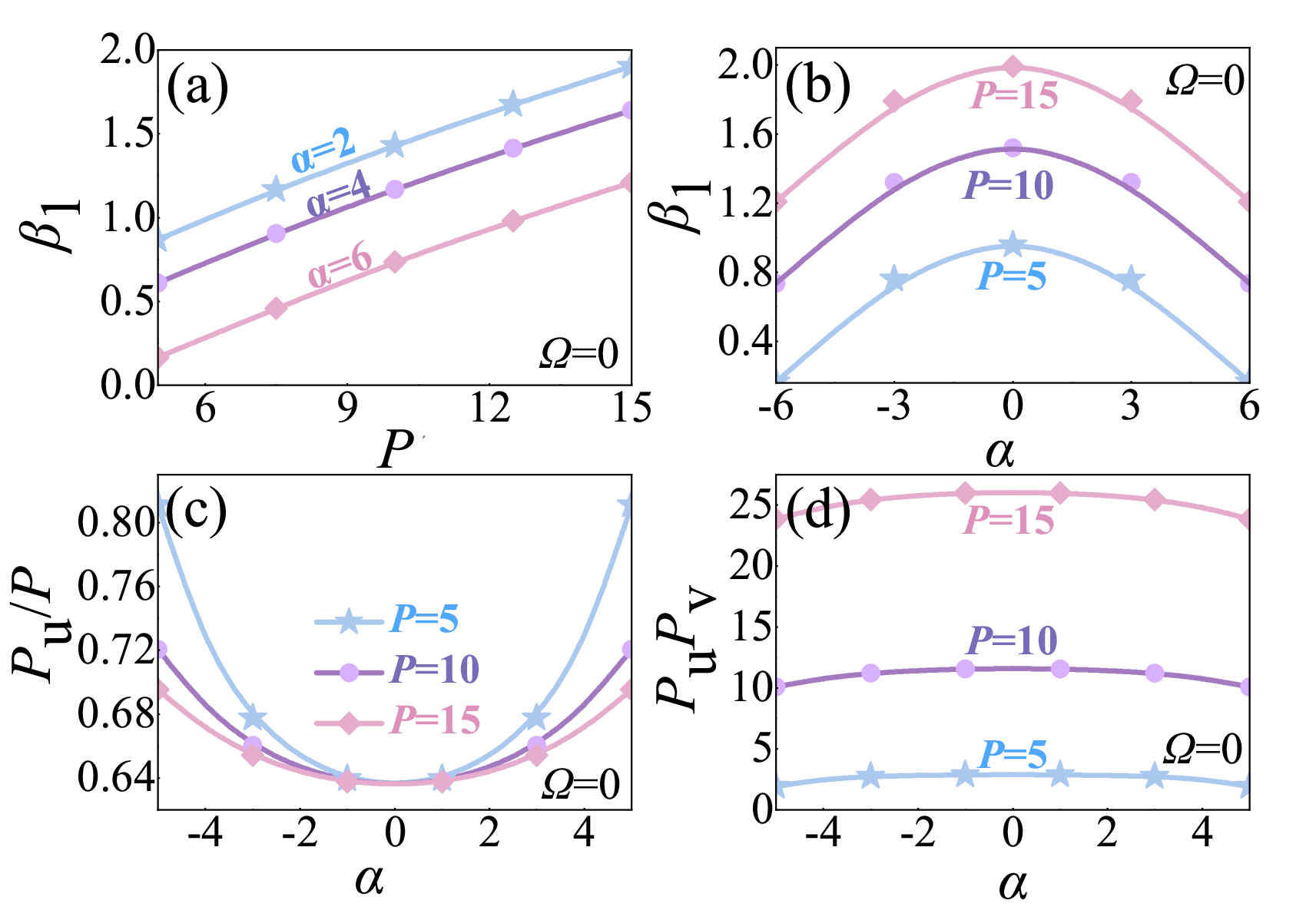}}
\caption{The propagation constant ($\beta_{1}$ [see Eq. (\ref{uzx})]) varies with $P$ and $\alpha$, and the $P_{u}/P$, $P_{u}P_{v}$ ([see Eq. (\ref{power})]) varies with $\alpha$. (a) $\beta_{1}$ with $P$ for different $\alpha$ ($\alpha$=2, 4, 6). (b) $\beta_{1}$ with $\alpha$ for different $P$ ($P$=5, 10, 15). (c) $P_{u}/P$ with $\alpha$ for different $P$ ($P$=5, 10, 15). (d) $P_{u}P_{v}$ with $\alpha$ for different $P$ ($P$=5, 10, 15). In these four graphs above, $\Omega=0$. ( In Fig. (a), the blue star represents $\alpha=2$, the purple circle represents $\alpha=4$, and the pink diamond represents $\alpha=6$. In Figs. (b,c,d), the blue star represents $P=5$, the purple circle represents $P=10$, and the pink diamond represents $P=15$.)}\label{beta}
\end{figure}

\begin{figure}[htbp]
\centering
{\includegraphics[scale=0.19]{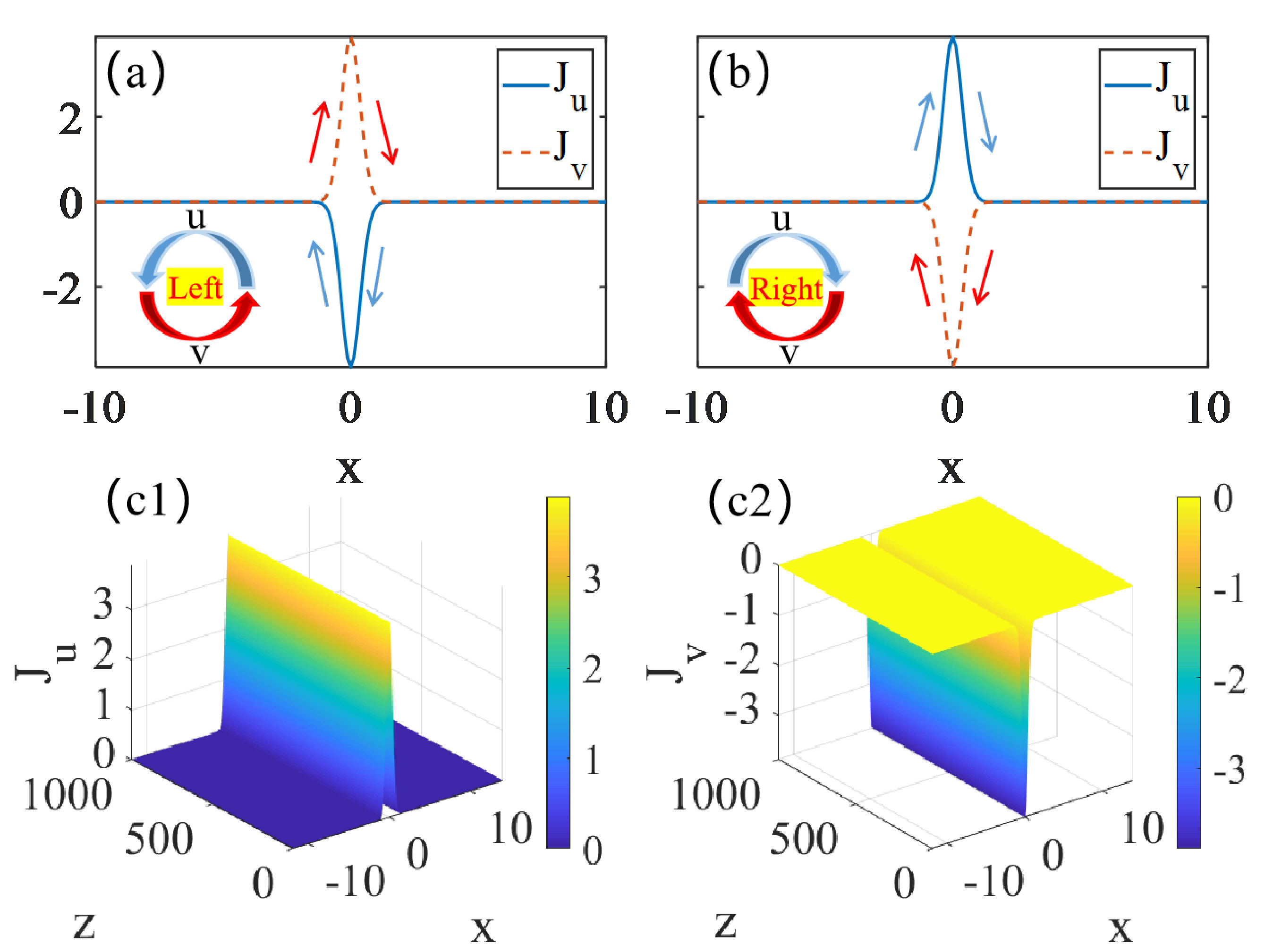}}
\caption{(a,b) The energy flow density diagram corresponding to the two cases. The blue solid curves and the red dashed curves are represent the energy flow density $J_{u}$ and $J_{v}$, respectively. (a) is the case of $\alpha>0$, (b) is the case of $\alpha<0$. The parameters are
($P$, $\Omega$, $\alpha$) = (10, 0, $\pm$4). (c1,c2) Direct simulations for energy flow density of the two components of the chiral soliton, at ($P$, $\Omega$, $\alpha$) = (10, 0, 4).}\label{RTP}
\end{figure}

Same result as our derivation, the coupled wave interacts with the equivalent magnetic field to produce chiral energy flow. Fig. \ref{RTP}(a) is the energy flow density of the FF and SH waves when $\alpha>0$, showing a circulating flow between them. And Fig. \ref{RTP}(b) is the case of $\alpha<0$. The arrows indicate the direction of energy flow. For the FF wave, when the energy flow density is less than 0, we define the direction of the flow of the energy flow as from right-to-left; when greater than 0, we define it as left-to-right. Accordingly, we complement the energy flow direction for the SH wave. The energy currents of the FF wave and the SH wave are converted into each other to form a ring current. As shown in Figs. 6(a,b), this results in two circulations that align with the left-hand and right-hand spiral rules. Figs. \ref{RTP}(c1,c2) demonstrates that the energy flow density of FF and SH waves are also stably transmitted in nonlinear photonic crystals.

\begin{figure}[h]
\centering
{\includegraphics[scale=0.295]{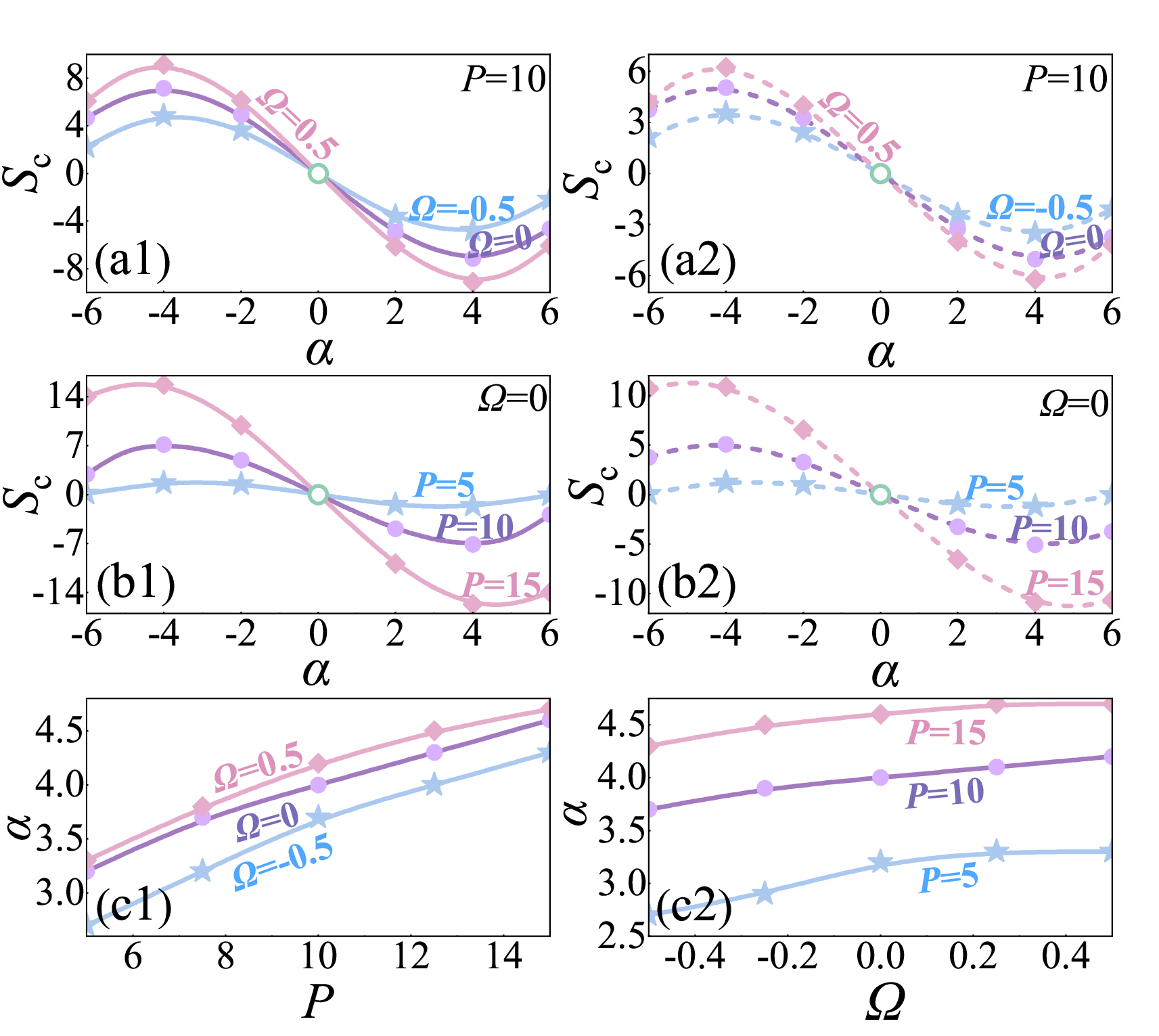}}
\caption{The variation in the chiral energy flow ($S_{c}$ [see Eq. (\ref{sc})]) and the $\alpha$(OI) corresponding to the "Optimal Inclination" with different parameters. (a1,b1) Numerical simulation results for $S_{c}$ with $\alpha$ for different $\Omega$ ($\Omega$=-0.5, 0, 0.5) and different $P$ ($P$=5, 10, 15); (a2,b2) Semi-analytical solution results for (a1,b1). (c1,c2) $\alpha$(OI) with $P$ and $\Omega$ for different other parameters. (In Figs. (a1,a2,c1), the blue star represents $\Omega=-0.5$, the purple circle represents $\Omega=0$, and the pink diamond represents $\Omega=0.5$. In Figs. (b1,b2,c2), the blue star represents $P=5$, the purple circle represents $P=10$, and the pink diamond represents $P=15$.)}\label{PP}
\end{figure}

Regarding the chiral flow $S_{c}$, there are two ways to calculate it. The first one is to simulate it numerically directly, and the second one is to compute it using our  semi-analytical solution, see Eq. (\ref{sc}). In order to further explore the relationship between the chiral energy flow $S_{c}$ of chiral solitons and each parameter, we analyzed the value of $S_{c}$ for different parameters. Figs. \ref{PP}(a1, b1)show the results of numerical simulation, the absolute value of the $S_{c}$ increases with the absolute value of the angular parameter $\alpha$ for a fixed power value of $P=10$ and a fixed $\Omega=0$. Figs. \ref{PP}(a2, b2) show the results of the semi-analytical solution, and it can be seen that the  semi-analytical solution results follow the same trend as the numerical simulation results, and at $\alpha=0$, $S_{c}=0$. When $\alpha$ increases to a certain value, the absolute value of the $S_{c}$ no longer increases and starts to decrease, i.e., there exists an "Optimal Inclination" that maximizes the $S_{c}$ for a fixed $P$ and $\Omega$, we refer to this point $\alpha$ as $\alpha$(OI). With Fig. \ref{beta} we know that the larger $|\alpha|$ is, the smaller $\beta P_{u}P_{v}$ is. Since the total power $P$ is a constant value, With Eq. (\ref{sc}) we can deduce that the presence of $\alpha$(OI) is the result of a compromise of $\alpha$ with propagation constant and the power of two waves. $S_{c}$ shows central symmetry for $\alpha<0$ and $\alpha>0$. When $\alpha=0$, the corresponding point is undefined, i.e., there is no chiral energy flow when there is no inclination in the positive and negative polarization directions of the crystal. Furthermore, we can see in Figs. \ref{PP}(a1,a2) that the larger the $\Omega$, the larger the absolute value of $S_{c}$ for a fixed $\alpha$ and $P$. In Figs. \ref{PP}(b1,b2), we can see that the larger the $P$, the larger the absolute value of $S_{c}$ for fixed $\alpha$ and $\Omega$. In Figs. \ref{PP}(c1,c2), we investigate the $\alpha$(OI) under different $P$ and $\Omega$, and the results show that the $\alpha$(OI) is positively correlated with the $P$ and $\Omega$, while other parameters are fixed.

\begin{figure}[htbp]
\centering
{\includegraphics[scale=0.3]{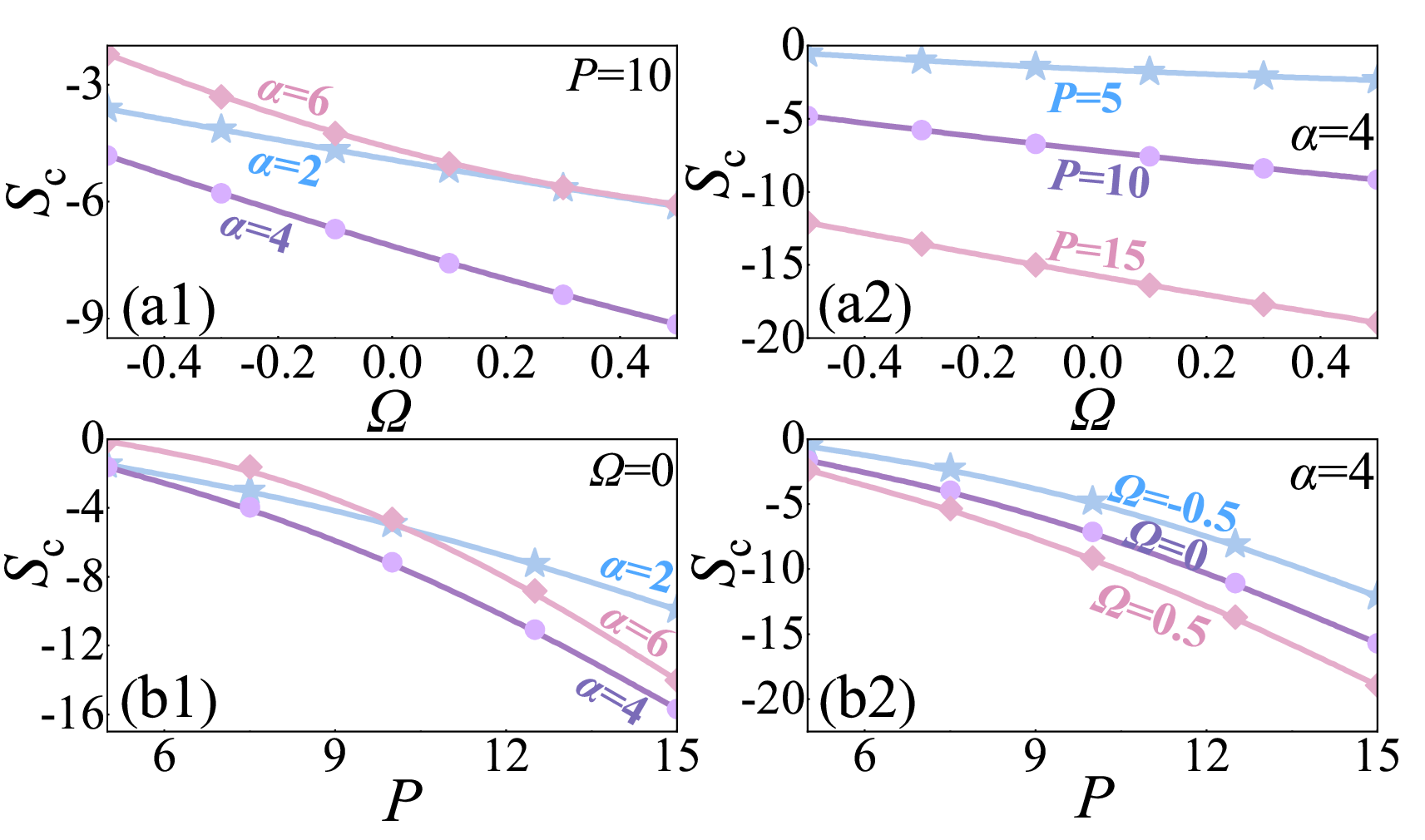}}
\caption{The variation in the chiral energy flow with different parameters. (a1,a2) $S_{c}$ with $\Omega$ for different $\alpha$ ($\alpha$=2, 4, 6) and different $P$ ($P$=5, 10, 15); (b1,b2) $S_{c}$ with $P$ for different $\alpha$ ($\alpha$=2, 4, 6) and different $\Omega$ ($\Omega$=-0.5, 0, 0.5). (In Figs. (a1,b1), the blue star represents $\alpha=2$, the purple circle represents $\alpha=4$, and the pink diamond represents $\alpha=6$. In Fig. (a2), the blue star represents $P=5$, the purple circle represents $P=10$, and the pink diamond represents $P=15$. In Fig. (b2), the blue star represents $\Omega=-0.5$, the purple circle represents $\Omega=0$, and the pink diamond represents $\Omega=0.5$. )}\label{OmegaP}
\end{figure}

In Figs. \ref{OmegaP}(a1, a2), the absolute value of $S_{c}$ increases with the increase of $\Omega$ for either fixed $P=10$ or fixed $\alpha=4$. In Fig. \ref{OmegaP}(a1), we can see that the absolute value of $S_{c}$ at $\alpha=4$ is the largest at different $\Omega$, which is because the $P$ is fixed to 10 at this time, and in Fig. \ref{PP}(c1), it can be seen that the $\alpha$(OI) value around 4 when $P=10$. And the $S_{c}$ curves corresponding to $\alpha=2$ and $\alpha=6$ cross, this is because when $\Omega=-0.5$, the numerical results obtained $\alpha$(OI) $=3.7$, and $\alpha=2$ is closer to this value, so $S_{c}$ is bigger, and with the increase of $\Omega$, the value of $\alpha$(OI) is also increasing, when $\Omega=0.5$, the value of $\alpha$(OI) is increases to 4.2, so at this point $\alpha=6$ approximates the $S_{c}$ corresponding to $\alpha=4$. In Figs. \ref{OmegaP}(b1,b2), the absolute value of $S_{c}$ increases with $P$ for either fixed $\Omega=0$ or fixed $\alpha=4$. In Fig. \ref{OmegaP}(b1), we can see that the absolute value of $S_{c}$ at $\alpha=4$ is the largest for different $P$. This is because the $\Omega$ is fixed to 0 at this time, and in Fig. \ref{PP}(c2), we can see that the value of $\alpha$(OI) is around 4 when $\Omega=0$. Here, the $S_{c}$ curves corresponding to $\alpha=2$ and $\alpha=6$ also cross, this is because when $P=5$, the numerical results obtained $\alpha$(OI) $=3.2$, and $\alpha=2$ is closer to this value, so the $S_{c}$ is larger, and with the increase of $P$, the value of $\alpha$(OI) is also increased, and when $P=15$, the value of $\alpha$(OI) is increases to 4.6, so at this point the $S_{c}$ corresponding to $\alpha=6$ is larger than the $S_{c}$ corresponding to $\alpha=2$. The results shown in Fig. \ref{PP} and Fig. \ref{OmegaP} validate each other.
\begin{figure}[h]
\centering
{\includegraphics[scale=0.23]{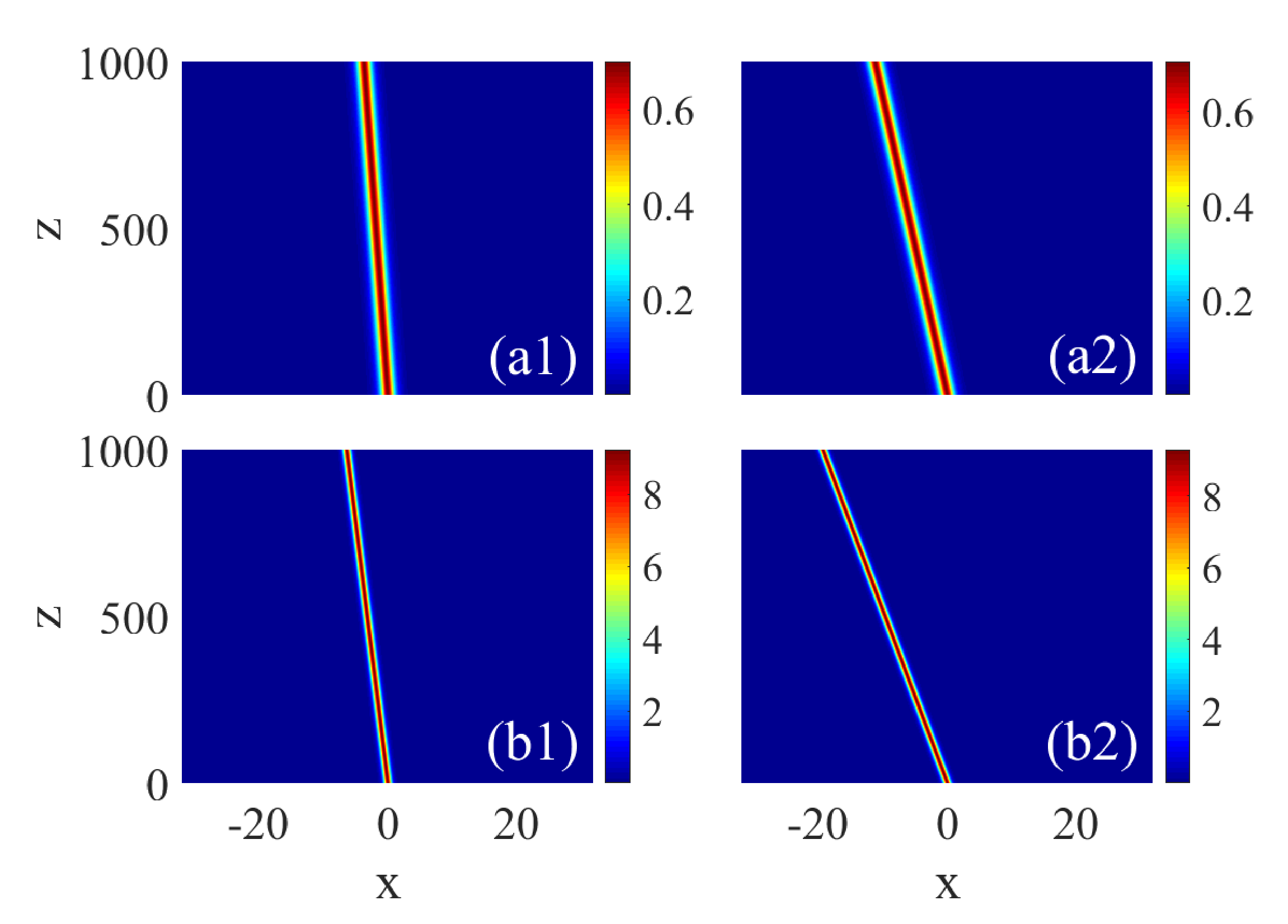}}
\caption{Dynamical behavior of chiral solitons under different kick strengths. (a1) Soliton is kicked by $K=0.01$ at the crystal with ($P$, $\alpha$, $\Omega$) = (2, 3, 0). (a2) ($K$, $P$, $\alpha$, $\Omega$) = ($0.03$, 2, 3, 0). (b1) ($K$, $P$, $\alpha$, $\Omega$) = ($0.01$, 15, 4, 0). (b2) ($K$, $P$, $\alpha$, $\Omega$) = ($0.03$, 15, 4, 0).}\label{kick}
\end{figure}
\section{Mobility and collision between the chiral soliton}
Here, we investigate the dynamical behavior of chiral solitons during transport in a quadratic nonlinear photonic crystal. It can be initialized by kicking solution as:
\begin{eqnarray}
&&\Psi _{u}(z=0)=\psi_{u}e^{iK x}, \label{ut}\\
&&\Psi _{v}(z=0)=\psi_{v}e^{i2K x}, \label{vt}
\end{eqnarray}
where $K$ is the strength of the kick. And collision between two moving chiral solitons is initialized with:
\begin{eqnarray}
\begin{aligned}
&\Psi _{u}(z=0)=\psi_{u}(x+x_{0})e^{iK x}+\psi_{u}(x-x_{0})e^{-iK x}, \label{ut}\\
&\Psi _{v}(z=0)=\psi_{v}(x+x_{0})e^{i2K x}+\psi_{v}(x-x_{0})e^{-i2K x}, \label{vt}
\end{aligned}
\end{eqnarray}
where $\psi_{u}(x\pm x_{0})$ represent the two initial stationary shapes of FF respectively, and $\psi_{v}(x\pm x_{0})$ represent SH. In this case, the expression for the symmetry of $\psi_{u}$ and $\psi_{v}$ satisfying the translation relation is as follows:
\begin{eqnarray}
&&\psi_{u}(x\pm x_{0})=\psi_{u}(x)e^{\pm i\frac{\alpha x_{0}}{3}}, \label{ut}\\
&&\psi_{v}(x\pm x_{0})=\psi_{v}(x)e^{\mp i\frac{\alpha x_{0}}{3}}, \label{vt}
\end{eqnarray}
thus, $2x_{0}$ is the distance between two colliding chiral solitons at $z=0$. Dynamics of the motion and collision of the chiral solitons can be numerically studied by launching above two condition to the direct simulation, respectively.

\begin{figure}[htbp]
\centering
{\includegraphics[scale=0.215]{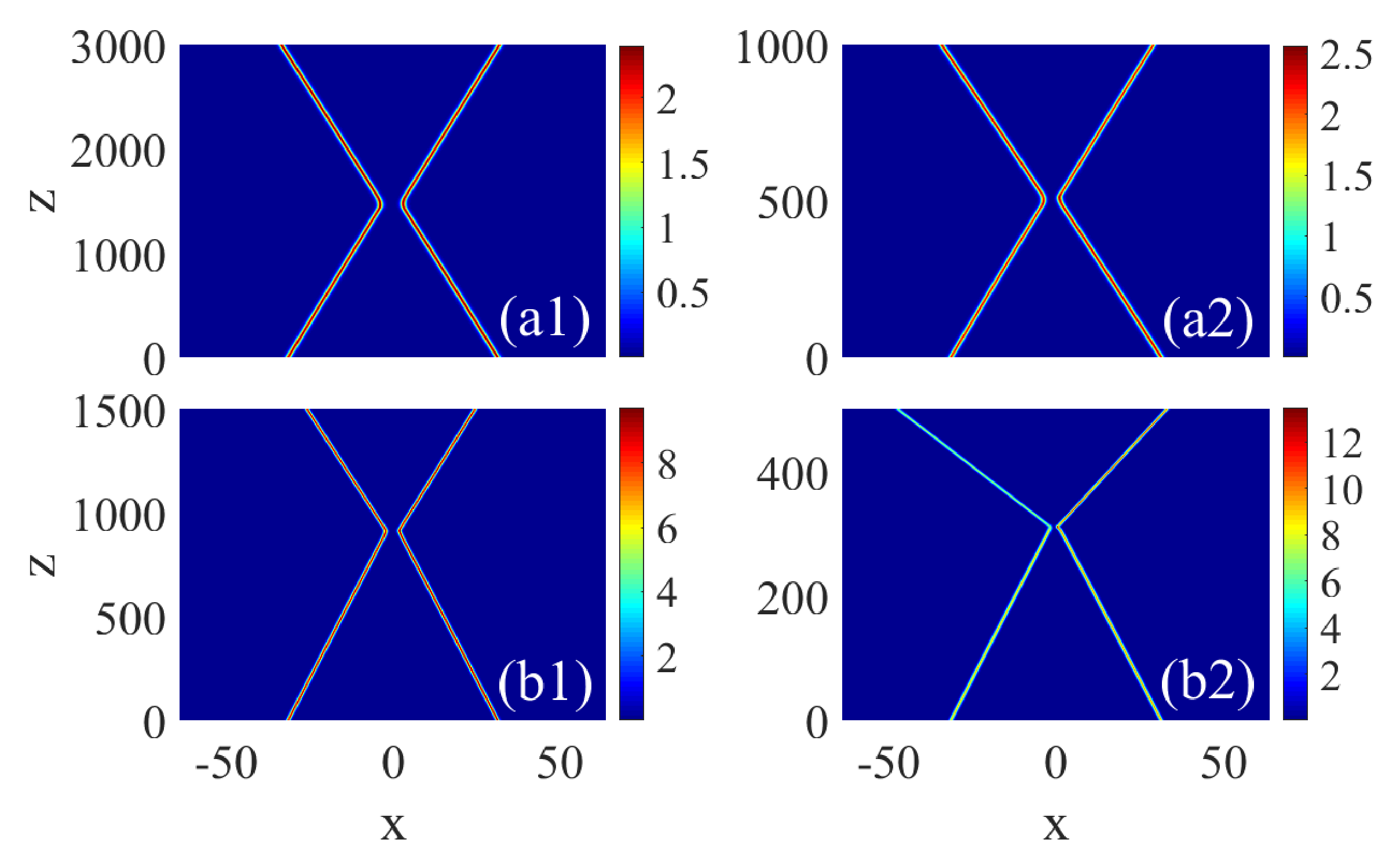}}
\caption{Dynamical behavior of chiral solitons under different collision strengths. (a1) Collision between two moving soliton kicked by $K=0.05$ on opposite direction at crystal with $(P, \alpha, \Omega) = (5, 4, 0)$. (a2) $(K, P, \alpha, \Omega) = (0.15, 5, 4, 0)$. (b1) $(K, P, \alpha, \Omega) = (0.5, 15, 4, 0)$. (b2) $(K, P, \alpha, \Omega) = (0.15, 15, 4, 0)$. }\label{collision}
\end{figure}

Numerical simulations show that the motion of the soliton can be induced by applying kicking power. As shown in Fig. \ref{kick}, the extent of the soliton transport trajectory shift increases with the increase of the kicking force, and the smaller the power, the smaller the soliton shift for the same kicking force. A typical example of a collision is shown in Fig. \ref{collision}, and it is interesting to note that a fully elastic collision between two chiral solitons is observed during the collision. We can see in Fig. \ref{collision}(b2) that when the kick increases, some energy escapes out of the soliton exterior during the collision, leading to an asymmetric elastic collision. The numerical results show that this energy escape is very small, on the order of $10^{-4}$, so it can be approximated as a fully elastic collision.

\section{ESTIMATION OF EXPERIMENTAL
PARAMETERS}
\begin{table}[h]
\begin{center}
\caption{Units of the dimensionless parameter.}

\label{tbl:bins}
\begin{tabular}{|cc|}
\hline
\multicolumn{1}{|c}{Dimensionless parameter} & \multicolumn{1}{c|}{Unit} \\
\hline
z=1                     & 0.0625 cm               \\
x=1                     & 7 $\mu$m                    \\
$|u|^2$=1, $|v|^2$=1          & 80 MW/cm$^{2}$, 160 MW/cm$^{2}$ \\
P=1                     & 40 W                    \\
$\Lambda$=1                     & 6.65 $\mu$m                   \\
\hline
\end{tabular}
\end{center}
\end{table}

The proposed 1D nonlinear photonic crystal is fabricated within a periodically poled lithium niobate crystal possessing a $\chi^{(2)}$ polarizability tensor factor $\chi^{(2)}$ of $d_0$=27 pm/V, $n_{1}\approx n_{2}\approx2.2$ and a thickness of 7 $\mu$m. We choose the fundamental frequency wave with a wavelength of 1064 nm and the corresponding second harmonic with a wavelength of 532 nm, and chose Type-1 polarization (the polarization directions for the two frequencies are identical), in $LiNbO_{3}$ crystals, $\Delta n\approx0.08$ \cite{DH1997}, when $\Lambda=6.65$ $\mu$m, the zero-phase matching condition can be achieved. An optical field amplitude $A_0$=200 kV/cm is uniformly applied within a $\chi^{(2)}$ nonlinear crystal. Utilizing Eq. (\ref{units}), the resulting characteristic distance $z_d$ is found to be 0.0625 cm. Table I provides the dimensionless covariates for the units involved in the analysis. The estimates in Table I indicate a transmission distance of $z$=1000, equivalent to a physical distance of 62.5 cm. This transmission distance of 62.5 cm is sufficient to empirically establish the stability of solitons within the system. And with $P=10$ and $\Omega=0$, the optimal inclination angle $\alpha$(OI) = 4, calculated from Eq. (\ref{units}) to be equivalent to the physical angle $\theta=30.96^{\circ}$, and the magnitude of the physical angle is also related to the optical field amplitude $A_0$.

In order to verify the feasibility of the current experimental setup, the generation of stable chiral solitons and chiral energy flow in a Gaussian beam as a natural input is simulated, which is initially needs to be loaded with a suitable phase using a spatial light modulator and is later projected into a nonlinear photonic crystal of length $z = 160$, and propagates only in the frequency shift (FF) component. The expression for the FF beam after loading the phase is:
\begin{eqnarray}
&&\Phi_{u}(z=0)=e^{-x^{2}/W^{2}}e^{-i\xi x}, \label{ue}
\end{eqnarray}
\begin{figure}[htbp]
\centering
{\includegraphics[scale=0.3]{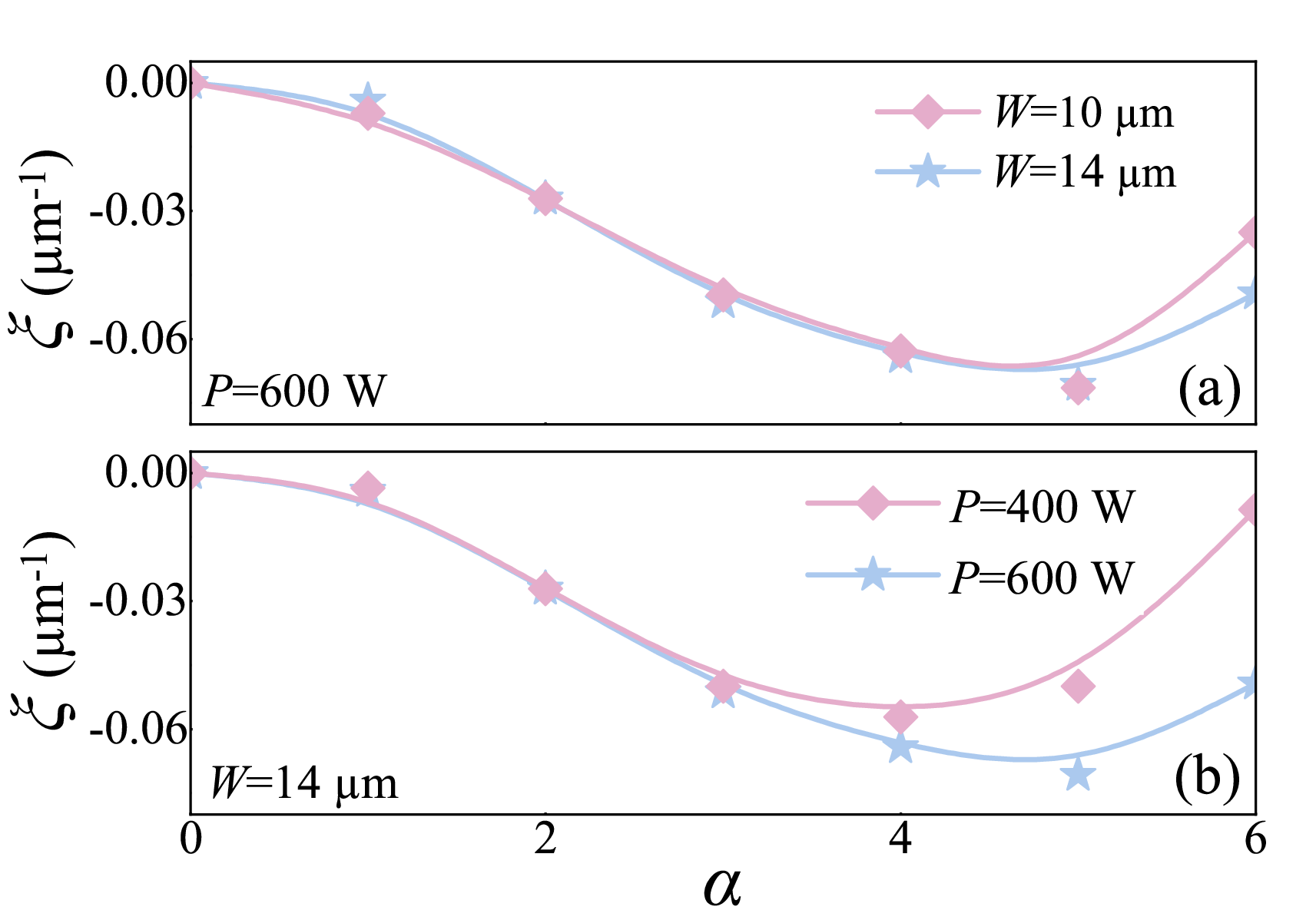}}
\caption{Variation of phase constant $\xi$ with $\alpha$ for different parameters. (a) $\xi$ with $\alpha$ for different $W$ ($W$ = 10 $\mu$m, 14 $\mu$m). (b) $\xi$ with $\alpha$ for different $P$ ($P$ = 400 W, 600 W). (In Fig. (a), the pink diamond curve and the blue star curve represents $W$ = 10 $\mu$m and $W$ = 14 $\mu$m, respectively. While in Fig. (b), these two curves represent $P$ = 400 W and $P$ = 600 W, respectively.)}\label{xixi}
\end{figure}where $W$ is the radius of the beam waist of the Gaussian beam and $\xi$ is the phase constant of the applied phase. The variation of $\xi$ with $\alpha$ for different $P$ and beam waist radii are shown in Fig. 11. In Fig. 11, we can see that $|\xi|$ increases and then decreases with $\alpha$ (which corresponds to inclination angle of the modulation stripes, $\theta$), and that there exists a value of $\alpha$ that maximizes $|\xi|$. This result agrees with the optimal inclination $\alpha$(OI) we obtained in Fig. \ref{PP}. Moreover, within a reasonable range of parameters, we can see in Fig. \ref{xixi} that the effects of $P$ and $W$ on $\xi$ are not significant.
\begin{figure}[htbp]
\centering
{\includegraphics[scale=0.177]{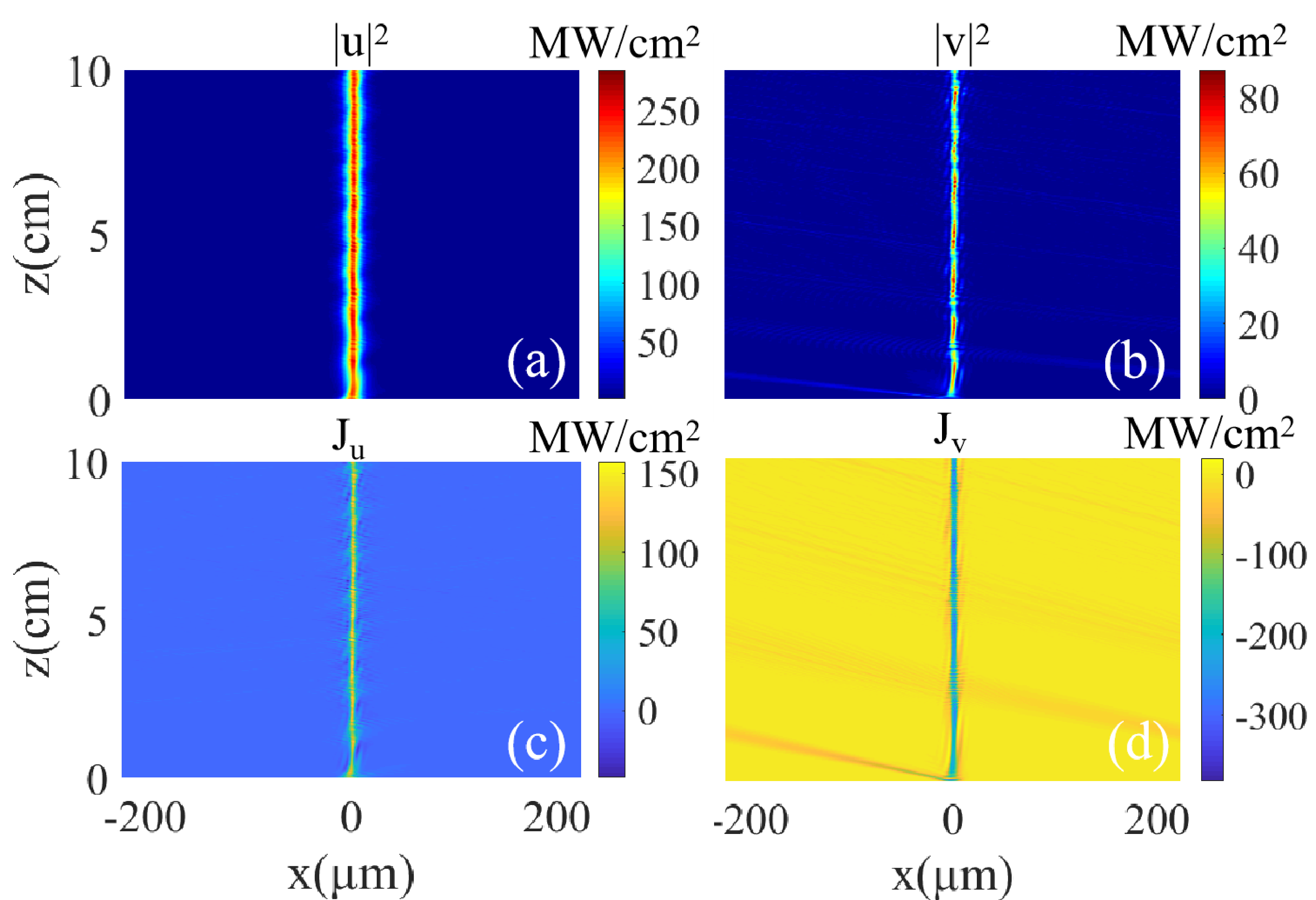}}
\caption{The simulated propagation of the input launched in
the FF component at $z=0$ in the form a Gaussian beam with $(P,\alpha,\Omega=15,6,0)$. (a,b) Transmission results for FF wave and SH wave of chiral solitons under natural evolution. (c,d) Transmission results for the energy flow density of chiral soliton FF wave and SH wave under natural evolution.}\label{evolution}
\end{figure}

Fig. \ref{evolution} shows a typical example demonstrating the results for chiral soliton transport of 10 cm in physical units. In the experiments, the length of nonlinear photonic crystals is usually in the range of 2 cm to 5 cm, and it is seen that the propagation distance of the Gaussian beam of the FF wave that naturally evolves to produce the chiral soliton and chiral energy flow can satisfy the experimental demand.

\section{Conclusion}
We introduced the quasi-phase-matched technique in nonlinear photonic crystals, and constructed an artificial gauge field, the equivalent magnetic field, by setting the inclination angle of the positive and negative polarization directions. Under the interaction between the equivalent magnetic field and the coupled wave, we found a chiral circulation with the direction of the energy flow conforming to the left-handed and right-handed helical rules. The solution of the chiral soliton is obtained based on convergence with the ITP method, the stability of the soliton solution is verified with the RTP method. At the same time, we derive the semi-analytical solutions for the chiral energy flow in this system. In the numerical simulations, we find that there exists a tilt angle in the polarization direction that maximizes the chiral energy flow, i.e., the "Optimal Inclination" (OI), and the chiral energy flow is positively correlated with the power and the detuning quantities, and the results of the semi-analytical solutions are corroborated with the results of the numerical simulations. Meanwhile, we probe the transport properties as well as the dynamical behavior of chiral solitons. The physical properties of the investigated chiral solitons are within the reach of existing experimental techniques. The findings in this paper provide new ideas for exploring the transport properties of solitons in $\chi^{(2)}$ nonlinear crystals.

\begin{acknowledgments}
We appreciate valuable discussions with Mr. Guilong Li, and Zibin Zhao,. This work was supported by NNSFC (China) through Grants No. 12274077, No. 12475014, the Natural Science Foundation of Guangdong province through Grant No. 2024A1515030131, No. 2023A1515010770, the Research Fund of GuangdongHong Kong-Macao Joint Laboratory for Intelligent Micro-Nano Optoelectronic Technology through grant No.2020B1212030010.
\end{acknowledgments}

\appendix
\section{Tuning of the detuning}
Here, we write the complete expression for the wave vector, where $k_{1,2}$ and represent the wave vectors of the fundamental frequency wave and the second harmonic, respectively. $n_{1,2}$ and $\lambda_{1,2}$ represent the refractive indices and wavelengths of the fundamental frequency wave and the second harmonic, respectively. Due to the different refractive indices of different wavelengths in the crystal, there will thus be a phase mismatch:
\begin{eqnarray}
\Delta k=2k_{1}-k_{2}=\frac{4\pi n_{1}}{\lambda_{1}}-\frac{2\pi n_{2}}{\lambda_{2}},\label{A1}
\end{eqnarray}
in the process of second harmonic generation, there exists $\lambda_{1}=2\lambda_{2}$. Therefore, we can write $\Delta k$ as:
\begin{eqnarray}
\Delta k=\frac{4\pi}{\lambda_{1}}(n_{1}-n_{2}).\label{A2}
\end{eqnarray}
In the preceding paragraphs, the expression for the amount of detuning is:
\begin{eqnarray}
{\Omega=z_{d}(\Delta k-\frac{2\pi}{\Lambda})}, \label{A3}
\end{eqnarray}
where $\Lambda$ is the inversion period of the QPM crystal. By periodically reversing the ferroelectric domain orientation in the material, the sign of the nonlinear coefficients is also periodically reversed, thus compensating for the phase mismatch \cite{RW2009}. Therefore, when $|\Delta k|=2\pi/\Lambda$, the detuning is 0 and phase matching can be achieved the condition of resonance can be reached.  At this point, there is:
\begin{eqnarray}
{\Lambda=\frac{1}{2}\frac{\lambda_{1}}{\Delta n}}, \label{A4}
\end{eqnarray}
where $\Delta n=|n_{1}-n_{2}|$.

In the Section.V of the paper, we selected $\lambda_{1}$=1064nm, $\lambda_{2}$=532nm, and chose Type-1 polarization, in $LiNbO_{3}$ crystals, $\Delta n\approx0.08$, when $\Lambda\approx6.65\mu$m, the zero-phase matching condition can be achieved. Then, if we want the detuning not to be 0, we only need to change the wavelength of the incident wave so that it is shifted from 1064nm.

\section{SCALED FORM OF THE DYNAMICAL COUPLED WAVE EQUATIONS FOR THE Second harmonic generation}
In Section. II, we give the following equations:
\begin{gather}
i\partial_{Z}A_{1}=-\frac{1}{2k_{1}}\partial_{X}^2A_{1}-\frac{2d(Z)\omega_{1}}{cn_{1}}A_{1}^{*}A_{2}e^{-i\Delta k Z}\label{B1},\\
i\partial_{Z}A_{2}=-\frac{1}{2k_{2}}\partial_{X}^2A_{2}-\frac{d(Z)\omega_{2}}{cn_{2}}A_{1}^{2}e^{i\Delta k Z}\label{B2},\\
d(Z)=d_0\mathrm{sgn}[\cos(2\pi Z/\Lambda +\varphi_{d})]\label{B3},\\
d(Z)=d_{0}\sum^{ }_{m\neq0}\frac{2}{\pi m}\sin(\frac{\pi m}{2})\exp[im(\frac{2\pi }{\Lambda}Z+\varphi_{d})],\label{B4}\\
\varphi_{d}=\frac{2\pi}{\Lambda}\Delta Z=\frac{2\pi}{\Lambda}\tan\theta X.\label{B5}
\end{gather}

In the following, we will show the later derivation in detail:

If we only select $m=\pm1$ from Eq. (\ref{B4}), which are assumed to provide best compensation for the phase mismatch, Eq. (\ref{B4}) can be simplified to
\begin{eqnarray}
\begin{aligned}
&d(Z)=d_{0}\frac{2}{\pi}\sin(\frac{\pi}{2})[\exp(i\frac{2\pi }{\Lambda}Z+\varphi_{d})\\
&\quad\quad\quad+\exp(-i(\frac{2\pi }{\Lambda}Z+\varphi_{d})],\label{B6}
\end{aligned}
\end{eqnarray}
the positive and negative frequency component of Eq. (\ref{B6}) can be substituted into Eqs. (\ref{B1},\ref{B2}), respectively (which is similar to the application of rotating wave approximation). Making a transformation of $A_{1,2}$ to a rotating frame $\tilde{A}_{1,2}$ by
\begin{eqnarray}
&&A_{1}=\tilde{A}_{1}\exp\left[-i\left( \Delta k-\frac{2\pi}{\Lambda}\right)Z\right],\\
&&A_{2}=\tilde{A}_{2}\exp\left[-i\left( \Delta k-\frac{2\pi}{\Lambda}\right)Z\right], \label{B7}
\end{eqnarray}
and substituting them into Eqs. (\ref{B1},\ref{B2}), the equations are changed to

\begin{align}
&i\partial_{Z}\tilde{A}_{1}=-\frac{1}{2k_{1}}\partial_{X}^2\tilde{A}_{1}-(\Delta k-\frac{2\pi}{\Lambda})\tilde{A}_{1}\notag\\
&\quad\quad\quad\quad-\frac{2d_{0}\omega_{1}}{cn_{1}}\frac{\pi}{2}\mathrm{sin}(\frac{\pi}{2})\tilde{A}_{1}^{*}\tilde{A}_{2}e^{i\varphi_{d}}\label{B8},\\
&i\partial_{Z}\tilde{A}_{2}=-\frac{1}{2k_{2}}\partial_{X}^2\tilde{A}_{2}-(\Delta k-\frac{2\pi}{\Lambda})\tilde{A}_{2}\notag\\
&\quad\quad\quad\quad-\frac{dd_{0}\omega_{2}}{cn_{2}}\frac{\pi}{2}\mathrm{sin}(\frac{\pi}{2})\tilde{A}_{1}^{2}e^{-i\varphi_{d}}\label{B9}.
\end{align}

By applying rescaling:

\begin{eqnarray}
&&\tilde{A}_{1}=u\sqrt{\frac{\omega_{1}I_{0}}{n_{1}}}, \quad \tilde{A}_{2}=v\sqrt{\frac{\omega_{2}I_{0}}{n_{2}}},\notag \\
&&I_{0}=(\frac{n_{1}}{\omega_{1}}+\frac{n_{2}}{\omega_{2}})|A_{0}|^{2},\quad
z^{-1}_{d}=\frac{2d_{0}\omega_{1}}{\pi cn_{1}}\sqrt{\frac{\omega_{2}}{n _{2}}I_{0}},\notag \\
&&z=Z/z_{d},\quad x=X\sqrt{k_{1}/z_{d}},\quad k_{1}/k_{2}\approx2,\notag \\
&&\quad n_{1}\approx n_{2}=n,\quad \alpha=\frac{2\pi}{\Lambda}\tan\theta\sqrt{\frac{z_{d}}{k_{1}}},\label{B10}
\end{eqnarray}
two dimensionless equations are then obtained:
\begin{eqnarray}
&&i\partial_{z}u=-\frac{1}{2}\ \frac{\partial^{2}u}{\partial x^{2}}\ - \Omega u- 2e^{i\alpha x}u^{*}v, \label{B12}\\
&&i\partial_{z}v=-\frac{1}{4}\ \frac{\partial^{2}v}{\partial x^{2}}\ - \Omega v- e^{-i\alpha x}u^{2}, \label{B13}
\end{eqnarray}
where $\alpha$ is a parameter related to the tilt angle $\theta$, and $\Omega$ is defined as Eq. (\ref{detuning}).


\begin{thebibliography}{99}
\bibitem{JS2004} J. Smith and G. Miller, Chiral solitons in nuclei: Electromagnetic form factors, Phys. Rev. C \textbf{70}, 065205 (2004).
\bibitem{AJ2020} A. J. Hess, G. Poy, J. S. B. Tai, S. \v{z}umer, and I. I. Smalyukh, Control of Light by Topological Solitons in Soft Chiral Birefringent Media, Phys. Rev. X \textbf{10}, 031042 (2020).








\bibitem{ME2013} M. Edmonds, M. Valiente, G. Juzeli¨±nas, L. Santos, and P. \"{O}hberg, Simulating an interacting gauge theory with ultracold bose gases, Phys. Rev. Lett. \textbf{110}, 085301 (2013).
\bibitem{JS2023} J. Song and Z. Yan, Formation, propagation, and excitation of matter solitons and rogue waves in chiral BECs with a current nonlinearity trapped in external potentials, Chaos: An Interdisciplinary Journal of Nonlinear Science \textbf{33}, 103132 (2023).
\bibitem{MM2010} M. Merkl, A. Jacob, F. Zimmer, P. \"{O}hberg, and L. Santos, Chiral confinement in quasirelativistic bose-einstein condensates, Phys. Rev. Lett. \textbf{104}, 073603 (2010).


\bibitem{VI2003} V. I. Kopp, Z. Zhang, and A. Z. Genack, Lasing in chiral photonic structures, Progress in Quantum Electronics \textbf{27}, 369 (2003).
\bibitem{YS2021} Y. Shen and I. Dierking, Electrically Driven Formation and Dynamics of Skyrmionic Solitons in Chiral Nematics, Phys. Rev. Applied \textbf{15}, 054023 (2021).



\bibitem{CZ2021} C. Zhang, J. Zhang, C. Liu, S. Zhang, Y. Yuan, P. Li, Y. Wen, Z. Jiang, B. Zhou, Y. Lei, D. Zheng, C. Song, Z. Hou, W. Mi, U. Schwingenschl\"{O}gl, A. Manchon, Z. Qiu, H. N. Alshareef, Y. Peng and X. Zhang, Chiral helimagnetism and one-dimensional magnetic solitons in a Cr-intercalated transition metal dichalcogenide, Advanced Materials \textbf{33}, 2101131 (2021).
\bibitem{SR2017} S. Riva, Chirality in metals: an asymmetrical journey among advanced functional materials, Materials Science and Technology \textbf{33}, 795-808 (2017).


\bibitem{RD2019} R. J. Dingwall and P. \"{O}hberg, Stability of matter-wave solitons in a density-dependent gauge theory, Phys. Rev. A \textbf{99}, 023609 (2019).
\bibitem{RD2018} R. J. Dingwall, M. J. Edmonds, J. L. Helm, B. A. Malomed, and P. \"{O}hberg, Nonintegrable dynamics of matter-wave solitons in a density-dependent gauge theory, New J. Phys. \textbf{20}, 043004 (2018).
\bibitem{MJ2015} M. J. Edmonds, M. Valiente, and P. \"{O}hberg, Elementary excitations of chiral Bose-Einstein condensates, Europhysics Lett. \textbf{110}, 36004 (2015).



\bibitem{UA1996} U. Aglietti, L. Griguolo, R. Jackiw, S. Pi, and D. Seminara, Anyons and chiral solitons on a line, Phys. Rev. Lett. \textbf{77}, 4406 (1996).
\bibitem{EH1998} E. Harikumar, C. N. Kumar, and M. Sivakumar, Chiral solitons in a current coupled Schr\"{O}dinger equation with self-interaction, Phys. Rev. D \textbf{58}, 107703 (1998).


\bibitem{SS2008} S. S. P. Parkin, M. Hayashi, L. Thomas, Magnetic domain-wall racetrack memory, Science \textbf{320}, 190 (2008).
\bibitem{FJ2010} F. Jonietz, S. M\"{u}hlbauer, C. Pfleiderer, A. Neubauer, W. M¨¹nzer, A. Bauer, T. Adams, R. Georgii, P. B\"{o}ni, R. A. Duine, K. Everschor, M. Garst, A. Rosch, Spin transfer torques in MnSi at ultralow current densities, Science \textbf{330}, 1648 (2010).


\bibitem{RG2022} R. Gao, X. Qiao, Y. Ma, Y. Jian, A. Zhang, and J. Xue, Chiral matter-wave soliton in a Bose-Einstein condensate under density-dependent gauge potential, Phys. Lett. A \textbf{446}, 128283 (2022).
\bibitem{KF2012} K. Fang, Z. Yu, and S. Fan, Photonic aharonov-bohm effect based on dynamic modulation, Phys. Rev. Lett. \textbf{108}, 153901 (2012).
\bibitem{WN2024} W. N. Faugno, M. Salerno, and T. Ozawa, Density dependent gauge field inducing emergent su-schrieffer-heeger physics, solitons, and condensates in a discrete nonlinear schr. \"{O}dinger Equation, Phys. Rev. Lett. \textbf{132}, 023401 (2024).



\bibitem{RO2011} R. O. Umucal{\i}lar and I. Carusotto, Artificial gauge field for photons in coupled cavity arrays, Phys. Rev. A \textbf{84}, 043804 (2011).
\bibitem{SL2013} S. Longhi, Effective magnetic fields for photons in waveguide and coupled resonator lattices, Opt. Lett. \textbf{38}, 3570-3573 (2013).
\bibitem{FL2015} F. Liu and J. Li, Gauge field optics with anisotropic media, Phys. Rev. Lett. \textbf{114}, 103902 (2015).
\bibitem{NS2016} N. Schine, A. Ryou, A. Gromov, A. Sommer, J. Simon, Synthetic Landau levels for photons, Nature \textbf{534}, 671-675 (2016).
\bibitem{SM2023} S. Ma, H. Jia, Y. Bi, S. Ning, F. Guan, H. Liu, C. Wang, and S. Zhang, Gauge Field Induced Chiral Zero Mode in Five-Dimensional Yang Monopole Metamaterials, Phys. Rev. Lett. \textbf{130}, 243801 (2023).














\bibitem{MA2014} M. Atala, M. Aidelsburger, M. Lohse, J. T. Barreiro, B. Paredes, and I. Bloch, Observation of chiral currents with ultracold atoms in bosonic ladders, Nature Phys. \textbf{10}, 588 (2014).
\bibitem{AC2014} A. Celi, P. Massignan, J. Ruseckas, N. Goldman, I. B. Spielman, G. Juzeli¨±nas, and M. Lewenstein, Synthetic Gauge Fields in Synthetic Dimensions, Phys. Rev. Lett. \textbf{112}, 043001 (2014).
\bibitem{DH2014} D. H\"{u}gel and B. Paredes, Chiral ladders and the edges of quantum Hall insulators, Phys. Rev. A \textbf{89}, 023619 (2014).


\bibitem{GL2023} G. Liu, X. Zhang, X. Zhang, Y. Hu, Z. Li, Z. Chen, and S. Fu, Spin-orbit Rabi oscillations in optically synthesized magnetic fields, Light Sci Appl \textbf{12}, 205 (2023).
\bibitem{GL2024} G. Liu, Z. Zeng, H. Lin, Y. Hu, Z. Li, Z. Chen, and S. Fu, Electrically engineering synthetic magnetic fields for polarized photons, Optica \textbf{11}, 980 (2024).

\bibitem{OY2022} O. Yesharim, A. Karnieli, S. Jackel, G. Di Domenico, S. Trajtenberg-Mills, and A. Arie, Observation of the all-optical Stern-Gerlach effect in nonlinear optics, Nat. Photon. \textbf{16}, 582 (2022).
\bibitem{AKa2018} A. Karnieli, and A. Arie, All-Optical Stern-Gerlach Effect, Phys. Rev. Lett. \textbf{120}, 053901 (2018).
\bibitem{Aviv2018}A. Karnieli, and A. Arie, Frequency domain Stern¨CGerlach effect for photonic qubits and qutrits, Optica \textbf{5}, 1297 (2018).
\bibitem{FZ2021} F. Zhao, J. L\"{u}, H. He, Y. Zhou, and Y. Li, Geometric phase with full-wedge and half-wedge rotation in nonlinear frequency conversion, Opt. Express \textbf{29}, 21820 (2021).


\bibitem{AK2022} A. Karnieli and A. Arie, The geometric phase in nonlinear frequency conversion, Front. Phys. \textbf{17}, 12301 (2022).
\bibitem{AK2018}A. Karnieli, and A. Arie, Fully controllable adiabatic geometric phase in nonlinear optics, Opt. Express \textbf{26}, 4920 (2018).
\bibitem{AK2019}A. Karnieli, S. Trajtenberg-Mills, G. D. Domenico, and A. Arie, Experimental observation of the geometric phase in nonlinear frequency conversion, Optica \textbf{6}, 1401 (2019).
\bibitem{YL2020} Y. Li, O. Yesharim, I. Hurvitz, A. Karnieli, S. Fu, and A. Arie, Adiabatic geometric phase in fully nonlinear three-wave mixing, Phys. Rev. A \textbf{101}, 033807 (2020).
\bibitem{YL2021} Y. Li, J. L\"{u}, S. Fu, and A. Arie, Geometric representation and the adiabatic geometric phase in four-wave mixing processes, Opt. Express \textbf{29}, 7288 (2021).

\bibitem{CE2000} C. Etrich, F. Lederer, B. A. Malomed, T. Peschel, and U. Peschel, Optical Solitons in Media with a Quadratic Nonlinearity, In Progress in Optics \textbf{41}, 483-568 (2000).
\bibitem{VE1999} V. E. Zakharov and S. Wabnitz, editors, Optical Solitons: Theoretical Challenges and Industrial Perspectives: Les Houches Workshop, September 28-October 2, 1998 (Springer Berlin Heidelberg, Berlin, Heidelberg, 1999).
\bibitem{BZ2016} B. Zhou and M. Bache, Invited Article: Multiple-octave spanning high-energy mid-IR supercontinuum generation in bulk quadratic nonlinear crystals, APL Photonics \textbf{1}, 050802 (2016).


\bibitem{JY2003} J. Yang and Z. H. Musslimani, Fundamental and vortex solitons in a two-dimensional optical lattice, Opt. Lett. \textbf{28}, 2094 (2003).
\bibitem{JY2004} J. Yang, I. Makasyuk, A. Bezryadina, and Z. Chen, Dipole and Quadrupole Solitons in Optically Induced Two-Dimensional Photonic Lattices: Theory and Experiment, Stud Appl Math \textbf{113}, 389 (2004).
\bibitem{JYI2004} J. Yang, I. Makasyuk, A. Bezryadina, and Z. Chen, Dipole solitons in optically induced two-dimensional photonic lattices, Opt. Lett. \textbf{29}, 1662 (2004).
\bibitem{YS2008} Y. Sivan, G. Fibich, B. Ilan, and M. I. Weinstein, Qualitative and quantitative analysis of stability and instability dynamics of positive lattice solitons, Phys. Rev. E \textbf{78}, 046602 (2008).



\bibitem{HS2023} H. Su, Y. Guo, Y. Guan, and H. He, The transport of dipole solitons in a one-dimensional nonlinear photonic crystal, Phys. Lett. A \textbf{478}, 128909 (2023).
\bibitem{MM1992} M. M. Fejer, G. A. Magel, D. H. Jundt, and R. L. Byer, Quasi-phase-matched second harmonic generation: tuning and tolerances, IEEE J. Quantum Electron. \textbf{28}, 2631 (1992).
\bibitem{JL2022} J. L\"{u}, B. Ma, and X. Wang, Quasi-phase-matching based on hilbert fractal superlattice structure, Chin. J. Laser \textbf{49}, 0608001 (2022).




\bibitem{FZ2023} F. Zhao, X. Xu, H. He, L. Zhang, Y. Zhou, Z. Chen, B. A. Malomed, and Y. Li, Vortex Solitons in Quasi-Phase-Matched Photonic Crystals, Phys. Rev. Lett. \textbf{130}, 157203 (2023).
\bibitem{AB2010} A. Bahabad, M. M. Murnane, and H. C. Kapteyn, Quasi-phase-matching of momentum and energy in nonlinear optical processes, Nature Photon \textbf{4}, 570 (2010).
\bibitem{DZ2008} D. Zhang, L. Qian, K. Wang, and H. Zhu, Influence of Cascaded Nonlinear Phase Shifts on Second-Harmonic Generation in High-Intensity Pumped QPM Structures, Chinese Phys. Lett. \textbf{25}, 3685 (2008).
\bibitem{KM1994} K. Mizuuchi, K. Yamamoto, M. Kato, and H. Sato, Broadening of the phase-matching bandwidth in quasi-phase-matched second-harmonic generation, IEEE J. Quantum Electron. \textbf{30}, 1596 (1994).
\bibitem{ZF2025} Z. Fan, W. Liu, L. Wang, W. Peng, D. Wu, S. Xu, and Y. Zhao, Vortex solitons in quasi-phase-matched photonic crystals with competing quadratic and cubic nonlinearity, Phys. Rev. E \textbf{111}, 034208 (2025).
\bibitem{CK2024} C. Kong, J. Li, X. Tang, X. Li, J. Jiao, J. Cao, and H. Deng, Composite solitary vortices of three-wave mixing in quasi-phase-matched photonic crystals, Chaos, Solitons $\&$ Fractals \textbf{187}, 115358 (2024).


\bibitem{BZ2025} B. Zhang, W. Yan, and F. Chen, Recent advances in femtosecond laser direct writing of three-dimensional periodic photonic structures in transparent materials, Adv. Photon. \textbf{7}, (2025).



\bibitem{AP2004} A. Picinin, M. H. Lente, J. A. Eiras, and J. P. Rino, Theoretical and experimental investigations of polarization switching in ferroelectric materials, Phys. Rev. B \textbf{69}, 064117 (2004).
\bibitem{SZ1997} S. Zhu, Y. Zhu, and N. Ming, Quasi-Phase-Matched Third-Harmonic Generation in a Quasi-Periodic Optical Superlattice, Science \textbf{278}, 843 (1997).
\bibitem{LT2011} L. Tian and X. Chen, Optical vortex converter with helical-periodically poled ferroelectric crystal, Opt. Express \textbf{19}, 11591 (2011).
\bibitem{AA2010} A. Arie and N. Voloch, Periodic, quasi-periodic, and random quadratic nonlinear photonic crystals, Laser \& Photonics Reviews \textbf{4}, 355 (2010).

\bibitem{YZ2021} Y. Zhang, Y. Sheng, S. Zhu, M. Xiao, and W. Krolikowski, Nonlinear photonic crystals: from 2D to 3D, Optica \textbf{8}, 372 (2021).

\bibitem{CB2002} C. Bowden and A. Zheltikov, Nonlinear optics of photonic crystals introduction, J. Opt. Soc. Am. B \textbf{19}, 2046 (2002).
\bibitem{MS2009} M. Skorobogatiy, Fundamentals of photonic crystal guiding, Cambridge University Press (2009).
\bibitem{AB2008} A. Bahabad and A. Arie, Engineering two-dimensional nonlinear photonic quasi-crystals, Opt. Lett. \textbf{33}, 1386 (2008).
\bibitem{VB1998} V. Berger, Nonlinear photonic crystals, Phys. Rev. Lett. \textbf{81}, 4136 (1998).



\bibitem{YG2014} Y. Guo, X. Xu, Z. Chen, Y. Zhou, B. Liu, H. He, Y. Li, and J. Xie, Three-Wave Mixing of Dipole Solitons in One-Dimensional Quasi-Phase-Matched Nonlinear Crystals, Chinese Phys. Lett. \textbf{41}, 014204 (2024).
\bibitem{SL2023} S. Liu, L. Wang, L. M. Mazur, K. Switkowski, B. Wang, F. Chen, A. Arie, W. Krolikowski, and Y. Sheng, Highly Efficient 3D Nonlinear Photonic Crystals in Ferroelectrics, Advanced Optical Materials \textbf{11}, 2300021 (2023).
\bibitem{YT2024} Y. Tang, K. Sripathy, H. Qin, Z. Lu, G. Guccione, J. Janousek, Y. Zhu, M.M. Hasan, Y. Iwasa, P.K. Lam, Y. Lu, Quasi-phase-matching enabled by van der Waals stacking, Nat Commun \textbf{15}, (2024).




\bibitem{RW2014} R. Wei and E. J. Mueller, Theory of bosons in two-leg ladders with large magnetic fields, Phys. Rev. A \textbf{89}, 063617 (2014).
\bibitem{QY2018} Q. Ye, X. Qin, Y. Li, H. Zhong, Y. S. Kivshar, and C. Lee, Band-gap structure and chiral discrete solitons in optical lattices with artificial gauge fields, Annals of Physics \textbf{388}, 173 (2018).



\bibitem{FY2008} F. Ye and L. Torner, Stabilization of dipole solitons in nonlocal nonlinear media, Phys. Rev. A \textbf{77}, 043821 (2008).
\bibitem{SS2006} S. Skupin, O Bang, and W. Krolikowsk, Stability of two-dimensional spatial solitons in nonlocal nonlinear media, Phys. Rev. E \textbf{73}, 066603 (2006).


\bibitem{KK2017} K. Krupa, K. Nithyanandan, U. Andral, and P. Grelu, Real-time observation of internal motion within ultrafast dissipative optical soliton molecules, Phys. Rev. Lett. \textbf{118}, 243901 (2017).
\bibitem{JF2003} J. Fleischer, T. Carmon, M. Segev, and D. Christodoulides, Observation of discrete solitons in optically induced real time waveguide arrays, Phys. Rev. Lett. \textbf{90}, 023902 (2003).






\bibitem{HS2014} H. Suchowski and A. Arie, Adiabatic processes in frequency conversion, Laser \& Photonics Rev. \textbf{8}, 333 (2014).

\bibitem{RW2009} R. W. Boyd and B. R. Masters, Nonlinear Optics, Third Edition, J. Biomed. Opt. \textbf{14}, 029902 (2009).

\bibitem{WM2009} W. Ma, Variational identities and applications to Hamiltonian structures of soliton equations, Nonlinear Anal.: Theory Methods \& Appl. \textbf{71}, e1716 (2009).
\bibitem{EA2019} E. Aslan, Optical soliton solutions of the NLSE with quadratic-cubic-Hamiltonian perturbations and modulation instability analysis, Optik \textbf{196}, 162661 (2019).
\bibitem{GP2013} G. Porat and A. Arie, Efficient, broadband, and robust frequency conversion by fully nonlinear adiabatic three-wave mixing, J. Opt. Soc. Am. B \textbf{30}, 1342 (2013).



\bibitem{DH1997} D. H. Jundt, Temperature-dependent Sellmeier equation for the index of refraction ne in congruent lithium niobate, Opt. Lett. \textbf{22}, 1553 (1997).



\bibitem{RA1997} R. A. Fuerst, M. T. G. Canva, D. Baboiu, and G. I. Stegeman, Properties of type II quadratic solitons excited by imbalanced fundamental waves, Opt. Lett. \textbf{22}, 1748 (1997).
\bibitem{AB2002} A. Buryak, Optical solitons due to quadratic nonlinearities: from basic physics to futuristic applications, Physics Reports \textbf{370}, 63 (2002).








\end{thebibliography}
\end{document}